\documentclass[preprint]{aastex61}
\usepackage[latin1]{inputenc}
\usepackage{amsmath}
\usepackage{amsfonts}
\usepackage{amssymb}
\usepackage{amsthm}
\usepackage{graphicx}
\usepackage{xcolor}
\usepackage{cancel}

\def\eg{\emph{e.g.},}

\def\etal{\emph{et al.~}}

\def\muG{~{\mu\rm G}}

\begin{document}

\title{Shocked Narrow-Angle Tail Radio Galaxies: Simulations and Emissions}

\author{Brian J. O'Neill}
\affiliation{School of Physics and Astronomy, University of Minnesota, Minneapolis, MN, USA}

\author{T. W. Jones}
\affiliation{School of Physics and Astronomy, University of Minnesota, Minneapolis, MN, USA}

\author{Chris Nolting}
\affiliation{School of Physics and Astronomy, University of Minnesota, Minneapolis, MN, USA}

\author{P. J. Mendygral}
\affiliation{Cray Inc., Bloomington, MN, USA}

\begin{abstract}

We present a numerical study of the interactions between the elongated AGN outflows representing an evolved, narrow-angle tail (NAT) radio galaxy and planar, transverse ICM shock fronts characteristic of those induced by galaxy cluster mergers (incident Mach numbers 2 - 4). The simulated NAT formation was reported previously in \cite{on19a}. Our simulations utilize a three-dimensional, Eulerian magnetohydrodynamic code along with energy-dependent Eulerian transport of passive cosmic ray electrons. Our analysis of the shock/NAT interaction applies a Riemann problem-based theoretical model to interpret complex shock front behavior during passage through the highly heterogeneous structures of the simulated NAT tails. In addition to shock compression, shock-induced vortical motions are observed within the tails that contribute to coherent turbulent dynamo processes that continue to amplify the magnetic fields in the tails well after initial shock compression. We analyze synthetic radio observations spanning the NAT-shock interaction period, and examine the brightness, spectral  and polarization properties of our shock-rejuvenated radio tails, as well as the extent to which the pre-shock states of the plasma and particle populations in our tails influence post-shock observations. Finally, we evaluate our findings in the possible context of a physical analogy to our simulated NAT providing the precursor to a cluster ``radio relic'' associated with an impacting ICM shock.

\end{abstract}

\section{Introduction} \label{sec:intro}
Radio galaxies (RG), synchrotron-luminous plasma outflows from active galactic nuclei (AGN), are common in galaxy clusters.
Radio luminous structures can extend up to multiple hundreds of kpc from the source AGN, interacting with the ambient intracluster medium  (ICM) as they develop. Accordingly, RGs can significantly impact the dynamics and thermodynamics of the ICM plasma.  For instance, centrally located RG are thought to engage in a feedback loop important to regulation of cooling of the ICM in non-merging cluster cores \citep[\eg][]{mn07}. At the same time RG dynamics and emissions are sometimes obviously strongly influenced by the structures and dynamics of their ICM environments. Consequently, RG properties, especially when spatially resolved, can reveal basic information about ICM structures and dynamical states. These complementary relationships underlie the case to understand the physical nature of RG/ICM interactions and their observable consequences.  

Clusters, as objects gravitationally collapsed within the cosmic web are, by their very nature, dynamically active objects. The infall of dark and baryonic matter onto clusters generates a dynamic environment in the ICM, driving winds, shocks, cold fronts, turbulence and sloshing motions to name a few consequences (e.g., \citealt{mv07,bru12,sch04,sim19}). Mpc-scaled shocks generated during mergers of similarly massed clusters (referred to as ``major mergers") are of particular interest, since their locations and strengths provide vital diagnostic information about the merging events (e.g., \citealt{mv07,vanw16,govoni19}). Since clusters are approximately virialized and the most massive structures in their part of the universe, relative velocities between comparably massed merging clusters are roughly similar to the associated cluster ICM sound speeds. Consequently, the Mach numbers, $\mathcal{M}$, of resulting merger shocks are modest. Cosmological simulations suggest more broadly that shocks with $2 \lesssim \mathcal{M} \lesssim 4$ play the dominant role in ICM heating (e.g., \citealp{ryu03,kang07}). 

Since ICM temperatures mostly fall in the range $\sim1$~-~$10$~keV, thermal plasma emissions are predominantly bremsstrahlung X-rays with associated, collisionally excited lines. Sections of ICM shock fronts are sometimes detected by way of discontinuities in projected X-ray emissions. But such edges are challenging to detect and interpret unless a shock element is projected edge on over a significant length. Mpc-scale cluster shocks, however, are generally curved, with nonuniform strengths, complicating their identification and characterization from X-ray jumps alone \citep[\eg][]{skill13,ha18}. Complimentary observations seem to be called for.

Other intrinsic shock identification tools include the thermal Sunyaev-Zeldovich (S-Z) effect \citep[\eg][]{mro19} and radio synchrotron emissions from cosmic ray electrons (CRe) accelerated at the shock. The latter of these in association with cluster merger shocks is the conventional explanation for so-called ``peripheral radio relics'' in clusters \citep[\eg][]{brun14}, and a number of radio relics have been identified with X-ray detected merger shocks \citep[\eg][and references therein]{vanw19}. On the other hand, some X-ray detected merger shocks do not appear to be associated with radio relics \citep[\eg][]{mark06,wilber18}. That realization and a growing body of plasma kinetic simulation work suggesting that possibly little or no injection of CRe from the thermal plasma takes place at ICM shocks \citep[\eg][]{ha18} has elevated the idea that significant radio emissions from merger shocks (so generation of radio relics) may depend on reacceleration of fossil CRe populations, rather than {\it{in situ}} injection and acceleration \citep[\eg][]{kang15,vanw17}. RG and their remnants are obvious potential candidates to supply such fossil CRe. Indeed, a number of  RG/ICM-shock interactions have been reported in the literature that support the idea that ICM shock encounters with RG lead to significantly enhanced radio emissions, possibly involving shock reacceleration of CRe \citep[\eg][]{bona14,shim15,vanw17}.

In an effort to understand the character of RG encounters with ICM shocks and the observable consequences, we have initiated an MHD simulation study of such interactions. The simulations include passive, but energy-dependent transport of CRe, so that resulting radio synchrotron properties can be modeled appropriately.  \citet{nolt19,nolt19b} explored interactions between ICM-strength shocks and lobed RGs. Here we extend that work to include such shock interactions with so-called ``narrow-angle tail'' (NAT) RGs. NATs, which are especially common in clusters, \citep[\eg][]{blan01,garon19} result from sustained relative motion between a RG and its ambient ICM; that is, the RG effectively evolves in a wind that deflects the RG jets and their discharge downwind. The NATs can extend multiple hundreds of kpc from their AGN sources, and, so long as the AGN jets remain ``powered on'' simulations suggest that NAT tails remain dynamic and highly inhomogeneous \citep{on19a}. Their considerable lengths and elongated morphologies make them interesting candidates to supply CRe at ICM shocks to form radio relics. We refer readers to \cite{on19a} and references therein for details of NAT formation dynamics. We do provide in \S \ref{sec:NAT} an outline of salient features of the simulated NAT presented in \citet{on19a} that are especially relevant to the present work.  Our focus in this paper is on the interactions of shocks with such a previously formed NAT. Specifically, we use new 3D MHD simulations to see what happens when plane shocks moving transverse to the NAT major axis collide with the NAT. That geometry is both relatively simple, and, on the face of it perhaps the best able to form a radio relic in such a manner. Our study includes cases with shock Mach numbers, $\mathcal{M} = 2, 3, 4$, in order to span the range of shock strengths most often mentioned in this context. As in the aforementioned studies we carry out synthetic, frequency-dependent radio synchrotron observations of our simulated objects in order to better understand the relationships between their dynamical and observable properties.

The rest of the paper is organized as follows: \S \ref{sec:NAT} outlines briefly especially relevant properties of the simulated NAT that provides the context for the simulations discussed in this paper. In \S~\ref{sec:SimDetails}, we outline the numerical simulations performed. In \S~\ref{sec:DynShk} we analyze the dynamics of the NAT-shock interaction, emphasizing the importance of the properties of the tails before shock impact. In \S~\ref{sec:SynObs}, we examine synthetic observations spanning each of the NAT-shock interactions. In \S \ref{sec:relics}, we discuss the implications of our results on the potential for NATs providing seed CRe for radio relics. We end with a summary of our conclusions in \S~\ref{sec:sum}.

\section{Key Properties of the Simulated Narrow-Angle Tail RG}\label{sec:NAT}
Narrow-Angle Tail RGs are examples of so-called ``tailed'' RGs in which the radio loud plasma jets emanating from the AGN  appear to be deflected towards each other into asymptotically almost parallel streams or tails. These tails can extend for hundreds of kpc. Often, but not always, both jets and both tails are distinguishable. The standard explanation for NATs is that transverse ram pressure due to relative motion between the AGN and the ambient medium (ICM) (a ``wind'') redirects the AGN jets downwind \citep{brb79,jo79} with a characteristic bending length, $\ell_b \sim \rho_w v_w^2/\rho_j v_j^2~r_j$, where $r_j$ is the radius of the jet in the cross-wind interaction region, while $\rho_w$ and $\rho_j$ are wind and jet mass densities respectively, and $v_w$ and $v_j$ are the wind and jet velocities, respectively. If the jets are in approximate pressure balance with the ambient wind ($P_j \approx P_w$), this can alternatively be expressed in terms of the internal jet and the wind Mach numbers as $\ell_b \sim (\mathcal{M}_j/\mathcal{M}_w)^2 ~r_j$ \citep{jnom17}. It is important to emphasize up front that $\ell_b$ is only a {\it{characteristic}} length, not a reliable, quantitative ``radius of curvature''. Quantitative, steady state jet trajectory models have been reported in previous works \citep[\eg][]{brb79}, including for arbitrary orientation between the wind and the initial jet axis \citep{on19a}. Those details are not central to the present work, but there are several NAT behaviors reported in \citet{on19a} that {\bf{are}} central to understanding the consequences of the impacts reported here between an existing NAT and planar ICM-strength shocks. 

To simplify our outline of the key NAT dynamical features we very briefly sketch the setup used in \cite{on19a} to simulate formation of the NAT referred to there as the ``reference simulation''. (More details follow in \S \ref{sec:SimDetails}.) The same 3D MHD code was used with the same grid, the same jet launching methods and the same jet and wind parameters as described below in \S \ref{sec:SimDetails} for the shocked NAT simulations. In short, bipolar, $\mathcal{M}_j =2.5$, magnetized, CRe-embedded jets with density $\rho_j = 0.01 \rho_w$ and $P_j \approx P_w$ were launched in the $\pm\hat{z}$ directions into a homogeneous, Mach 0.9, unmagnetized, CRe-free wind blowing orthogonal to the jets in the $+\hat{x}$ direction. The emergent jets carried a torroidal magnetic field with peak intensity, near the perimeter, $B_0=1 \muG$ (plasma $\beta_{p,j} = 25$). 

The jets were launched out of a stationary cylinder with effective radius, $r_j \approx 4.5$ kpc (9 grid cells, including a cylindrical, transition collar described in \S \ref{sec:SimDetails}). Thus, the nominal, characteristic bending length, $\ell_b \sim 35$ kpc. The full, transverse span of the two tails eventually was in some places $\sim 200$ kpc, so several times larger than $\ell_b$. The tail lengths were several times longer (see figures \ref{fig:BnC165}, \ref{fig:N3150} below), so that the ``NAT'' moniker is appropriate, despite the large scales. An objective in designing this study was to create at high resolution NAT structures with extents at least 1/2 Mpc. On the other hand, we emphasized in \cite{on19a} that all the dynamical lengths in that simulation (and, therefore, those reported here), including $\ell_b$, could be rescaled to larger or smaller values simply by resetting the value for $r_j$. If the velocities (\eg ~$v_j$ and $v_w$) remain unchanged, associated dynamical times would similarly scale with $r_j$. For example, if $r_j \sim 1$ kpc, the dynamical lengths and times reported here would both decrease by factors $\approx 4.5$.  On the other hand, the AGN jets also included CRe, which were tracked using energy dependent transport physics incorporating radiative losses from synchrotron and inverse Compton scattering of the CMB, here for redshift, z = 0.2. Those radiative losses are not scale invariant. The simulations require specific choices for the related parameters; namely magnetic field strengths and redshift.

Three dynamical characteristics of the simulated NAT stand together as keys to the subsequent shock interactions reported below. In short, the tails were highly heterogeneous in density, velocity and magnetic field before the arrival of the ICM shock. The first specific component leading to this character development was that the jets, as they were deflected downwind, were not in a steady equilibrium, even though they were steady at their source. Instead, as they bent, they tended to abruptly ``flap'' and sometimes, within what \cite{on19a} called the ``disruption zone'', pinch off where the local curvature became sharp, before resuming their drive towards the downwind direction (into the tails). The active jets and the pinched-off jet segments, which we call ``jet clumps'', provided fresh supplies of CRe and magnetic fields from the AGN well down the tails.
Consequently, the plasma in the tails was predominantly in two ``phases" (see the left panel of figure \ref{fig:BnC165}). The jet clumps, which were low density and carried concentrations of CRe, were surrounded by a medium that was a mixture of varying proportions of jet and ICM plasma (the ``diffuse" region). As the jet clumps traveled downstream, they gradually, but nonuniformly mixed with ICM plasma. The ``diffuse'' tail plasma tended to dominate the interiors of the two tails, while detached jet clumps maintained some of the $\hat{z}$ component momentum present as they detached, which carried them towards the transverse bounds of the tails. We note at the same time that all the $\hat x$ component (wind-aligned) momentum in the jets was necessarily extracted from the wind and concentrated into the jets as they were deflected. Subsequently, most of that momentum was redeposited broadly into the tails, but primarily into regions of lower density than the undisturbed wind. Consequently, the tails actually extended (lengthened) downwind faster than the wind itself. In this simulation the time averaged tail extension rate was $\sim 1.1~v_w$.

\begin{figure}[ht]
\centering
\includegraphics[width=\textwidth]{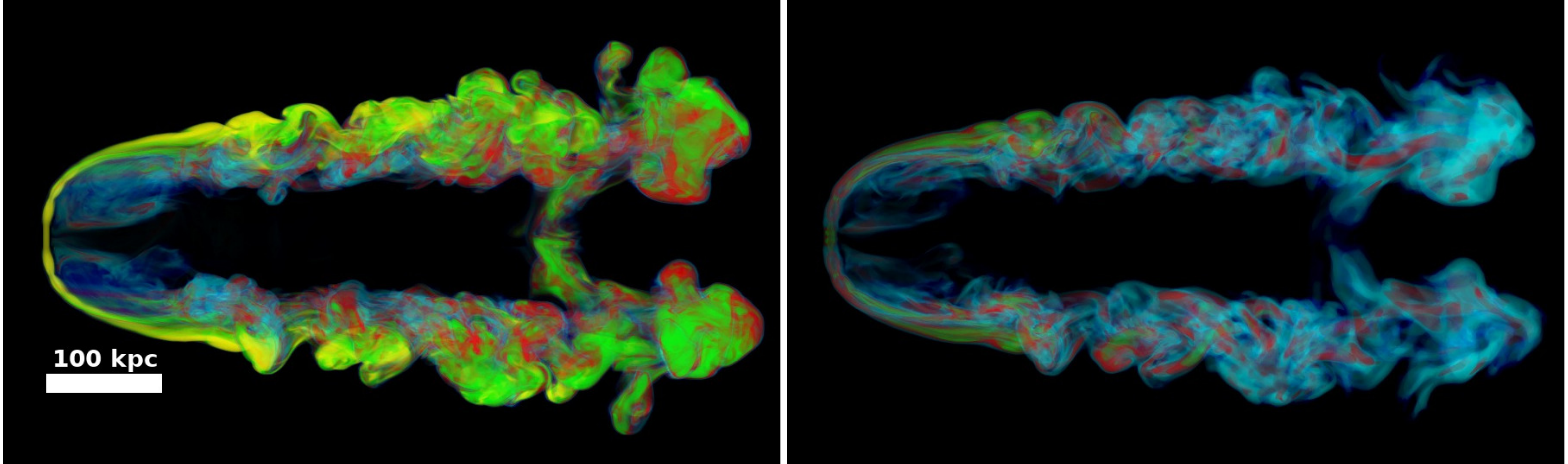}
\caption{Volume renderings of the NAT before shock impact at $t \approx 541$~Myr (viewed down the $\hat{y}$ axis). Left: Jet mass fraction, $C_j$. $C_j = 1$ for pure jet plasma; $C_j = 0$ for pure ICM plasma (yellow: $C_j > 0.8$; green: $0.6 < C_j < 0.8$;  red: $0.4 < C_j < 0.6$;  aqua: $0.2 < C_j < 0.4$;  blue: $C_j < 0.2$). Right: Magnetic field strength, $B$ (yellow: $B > 2\muG$; green: $1\muG < B < 2\muG$; red: $0.5\muG < B < 1\muG$; aqua: $0.2\muG < B < 0.5\muG$; blue: $B < 0.2\muG$). Data from run \textbf{N3} in Table \ref{tab:param}.  }
\label{fig:BnC165}
\end{figure}

The second key NAT dynamical feature, which was a consequence of the first, was that, although the incident wind in the simulation was quiet and homogeneous, the tails and the regions between the tails became turbulent, with outer turbulence scales, $\sim \ell_b$, and $\delta v \la 1/2 v_w$, leading very roughly to turbulence timescales, $t_{turb} \sim \ell_b/v_w \sim 100$ Myr. We note that these eddy velocities are large enough to influence the propagation of the ICM shock when it impacts, and that this will have significant consequences. It is important to keep in mind at the same time that the turbulence properties within the tails were neither homogeneous, nor steady. Of course, interactions with an initially turbulent wind \citep{porter09} would enhance these turbulent behaviors. 

\begin{figure}[ht]
\centering
\includegraphics[width=\textwidth]{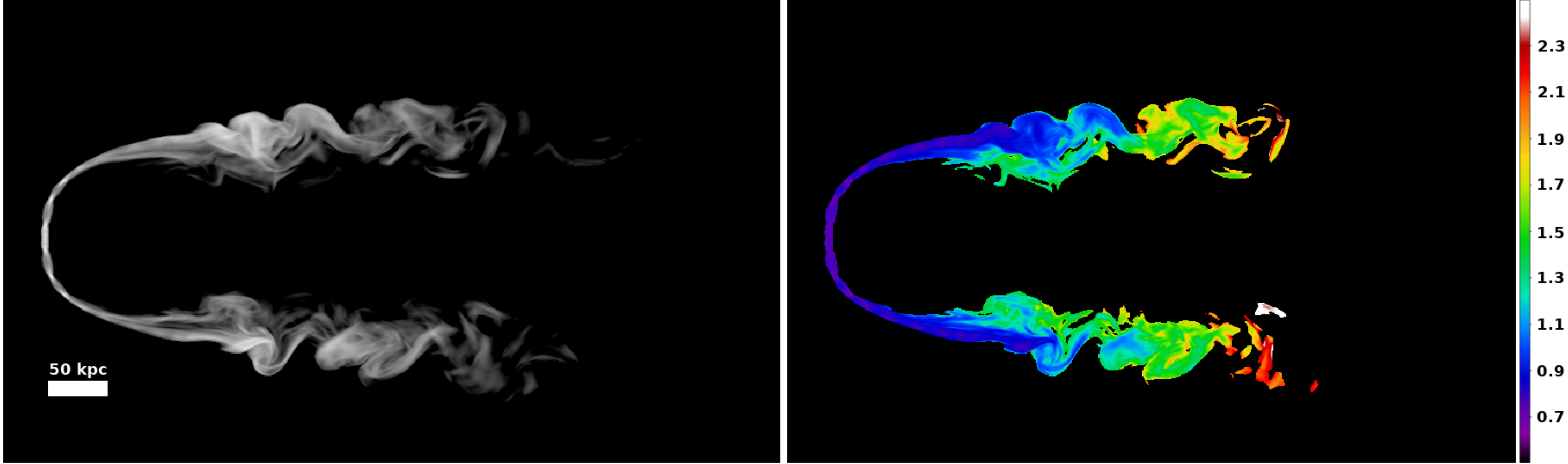}
\caption{Synthetic radio observations of the NAT structures illustrated in Figure \ref{fig:BnC165}. Left: 150 MHz synchrotron image, $I_{\nu,150}$, (arbitrary log scale; dynamical range 500:1). Right: Spectral index map, $\alpha_{150,325}$, for pixels with $I_{\nu,325} > (1/500) \max ( I_{\nu,325} )$. The observer's LoS  projects along the $+\hat{y}$ axis.}
\label{fig:N3150}
\end{figure}

The third key dynamical feature of the NAT formation that will especially influence synchrotron emissions from the shocked NAT was the development of an amplified, but highly intermittent and filamentary magnetic field distribution, particularly in the tails. This magnetic field topology was an outcome of the first two features, and is evident in the right panel of figure \ref{fig:BnC165} and in the synchrotron intensity image in figure \ref{fig:N3150}. Note, as well, that in the older tail sections (to the right in figure \ref{fig:BnC165}) the magnetic field strength had begun to decay as a result of decay of tail turbulence downwind and magnetic dissipation and diffusion (the numerically-based viscous and resistive dissipation scales in these simulations $\sim 1-2$ kpc). These fields will subsequently be ``rejuvinated'' by shock interaction, but at this stage their decay contributed substantially to reduced synchrotron visibility of the older tail sections.

The highly heterogeneous properties of the pre-shock NAT are evident in the synchrotron image shown in figure \ref{fig:N3150} from the same time as figure \ref{fig:BnC165}. The left panel shows the 150 MHz synchrotron brightness distribution in log scale, while the right panel shows the distribution of the radio spectral index between 150 MHz and 325 MHz, $\alpha_{150,325}$\footnote{The radio spectral index, $\alpha$ is defined as $\alpha_{\nu_1,\nu_2} = (log(I_1/I_2)/(log(\nu_2/\nu_1))$}. The older tail sections are virtually invisible by this time, both because of the aforementioned magnetic field decay and also due to radiative  cooling (``aging'') of the CRe. The filamentary character of the magnetic field is  obvious from the intensity image.

\section{New Simulation Details} \label{sec:SimDetails}
As already mentioned, the simulations presented here used the same code, the same grid, the same jet launching device and the same ICM wind as for the ``reference NAT'' simulation reported in \citet{on19a}. Indeed, that NAT provides the initial conditions for the new simulation study. We present here a summary of the most pertinent computational arrangements, referring readers to \citet{on19a} and references therein for full details.
\subsection{Basic Setup}\label{subsec:setup}

The simulations were carried out with  the non-relativistic 3D ``WOMBAT'' ideal, compressible MHD code  described in \cite{mendthesis}, utilizing  a Roe-based, 2nd order accurate ``TVD'' scheme along with the ``constrained transport"  (CT) method  to maintain a divergence-free magnetic field. We used a Cartesian, Eulerian grid with cells, $\Delta x = 0.5$ kpc, on a domain  spanning $x,y,z =\pm 513, \pm 121.5, \pm 283.5$~kpc (so $2052\times486\times1134$ cells). The domain was initialized with a uniform, adiabatic, $\gamma = 5/3$, unmagnetized ICM (the ``wind'') with density, $\rho_w = 5/3\times 10^{-28}~\rm{gm}~\rm{cm}^{-3}$, pressure, $P_w = 10^{-12}~\rm{dyne}~\rm{cm}^{-2}$ and velocity, $v_w = 900~\rm{km}~\rm{s}^{-1}$ in the $\hat{x}$ direction, which corresponds to a wind Mach number, $\mathcal{M}_w = 0.9$. The wind was maintained by continuous boundary conditions at the $-x$ domain boundary. Although real ICMs are neither unmagnetized, nor homogeneous on these scales, eliminating those complications in this initial study allowed us to better focus on basic physical behaviors.

A bidirectional, steady and supersonic light jet pair aligned with the $\hat{z}$ axis (so orthogional to the ICM wind) was initialized at the start of the simulation, $t = 0$, reported in \cite{on19a} and maintained during NAT formation. It was switched off during the extended simulations reported here at first shock encounter with the NAT as listed in Table \ref{tab:param} for each of three shock scenarios. The jets were launched from a stationary cylinder centered at position, $(-390,0,0)$ kpc.  So long as the jets were active, plasma conditions ($\gamma = 5/3$) were maintained in pressure equilibrium with the surroundings ($P_j \approx P_w$). Collimated  jet flows emerged from each launch cylinder end with density, $\rho_j = 0.01 \rho_w$ and velocity, $v_j = 2.5\times 10^4$~km s$^{-1}$, giving an emergent jet internal Mach number, $\mathcal{M}_j = 2.5$ and, when integrated across the jet cylinder cross section, a total jet kinetic power a bit under $10^{43}~\rm{erg/s}$.

A toroidal magnetic field, $B(r) = B_0(r/r_{jc})$, was maintained inside the active launch cylinder. The magnetic field strength at $r = r_{jc}$, $B_0 = 1~\mu$G, corresponded to a fiducial $\beta_{pj} = 8\pi P_j/B_0^2 = 25$. \cite{on19a} point out that, despite the fact that $\beta_{pj}\gg 1$ at launch, magnetic field amplification due especially to ``field line stretching'' creates substantial regions within the NAT structures where magnetic (``Maxwell'') stresses are dynamically important. We will see below, in \S \ref{sec:ShkMag} that enhanced turbulence during shock impact on the tails can further amplify those magnetic fields.  

We mention that such parameters as $\mathcal{M}_j$, $\rho_j/\rho_w$, $r_j$ and $B_0$ are not tightly constrained by theory or observation, so, to a large degree are adjustable parameters in our simulations. Our chosen values aimed at realistic, if still idealized, behaviors, while being practical in terms of the resulting computational requirements. 

Jet plasma emerging from the jet launch cylinder was marked, or ``colored'', by a passively advected, mass-weighted scalar field, $C_j =1$. Outside the jet launch cylinder, inititially, $C_j = 0$. So, throughout the simulations, any computational cell containing only AGN jet plasma took the value, $C_j = 1$, while any cell containing no AGN plasma had $C_j = 0,$ In cells developing a  mix of jet and ambient fluid $0 < C_j <1$. Thus, $C_j$ represents the fraction of the mass in a given cell at a given time that originated in the jet launch cylinder. Because $C_j$ behaves much like a pigment, we sometimes refer to it as ``color''.

Passive, relativistic, cosmic ray electrons (CRe) embedded in the  AGN fluid emerging from the jet launch cylinder were transported in space and momentum ($p \approx E/c = \Gamma_e m_e c$) using an Eulerian version of the  ``coarse grained momentum volume" (CGMV) transport algorithm \citep{jk05} for the CRe  distribution function, $f(\vec{r}, p, t)$, in the range $5~\rm{MeV}\le \Gamma_em_ec^2= pc\le 85~\rm{GeV}$, $10 \la \Gamma_e \la 1.7\times 10^5$.  The CRe momentum (energy) range was spanned by 8 uniform bins in $\ln{p}$.  At launch, $f(p) \propto p^{-q_o}$, with $q_o = 4.5$\footnote{The corresponding synchrotron spectral slope, $\alpha_o = (q_o-3)/2 = 0.75$}.  CRe adiabatic energy gains and losses were tracked, along with  synchrotron and inverse Compton (iC)  radiative losses off the CMB. 

CRe passing through shocks were subjected to standard test-particle diffusive shock (re)acceleration (DSA), for which the equilibrium electron distribution immediately downstream of the shock is a power law, $f\propto p^{-q_s}$, with slope  $q_s = 4 M_s^2/(M_s^2 - 1)$. Injection at shocks of thermal electrons into the CRe population was not included. At CRe energies of primary interest and environments modeled here Coulomb scattering energy losses are negligible compared to radiative losses. So, those losses were neglected. We did not include 2$^{nd}$ order Fermi (FII) turbulent particle acceleration, although turbulent flows do develop in the simulated scenarios, and FII acceleration could result. On the other hand, the net efficiency of FII acceleration is highly model dependent, so predictions are very uncertain. Since the simulated CRe were passive, we considered their number arbitrary. All their ``observed'' emissions are consequently presented in arbitrary units. We note, as well, because the undisturbed ICM contained no CRe\footnote{A tiny CRe population was included in the undisturbed ICM in order to avoid numerical singularities.} and was unmagnetized, only plasma with $C_j > 0$ contributed to synchrotron emissions.

\subsection{The ICM Shocks}\label{subsec:shockdefs}

Very roughly when the NAT reached longitudinal extent $\sim500$ kpc, we restarted the original simulation at times listed in Table \ref{tab:param}, and explained shortly, to follow its extension under two different conditions. One extension simply continued the evolution of the NAT to later times, with the AGN deactivated after a period of time detailed below. The other restarted simulations initialized at the restart time a steady, plane shock with Mach number, $\mathcal{M} =$ 2, 3 or 4 propagating from the $-\hat{z}$ boundary. The resulting shock planes were normal to the plane containing the two emergent jets, but aligned with the long axis of the NAT. Consequently, the shocks sequentially encountered each of tails transversely. Since our primary study objective was to examine shock interaction with the radio tails rather than the AGN itself, AGN activity was switched off at first contact between the shock and the NAT for the simulations reported here.  Each of the simulations then continued until after the shock completed passage through both tails. The restart/extension times were chosen so that each of the shock encounters was completed at approximately the same simulation time. Those shock simulations are labeled according to shock Mach number as M2, M3 and M4 in Table \ref{tab:param}. The simple wind extensions (without shocks) from the corresponding start times are labeled N2, N3 and N4 in Table \ref{tab:param}. Each continued  until the shock in the paired simulation, M2, M3 or M4, had completed its passage through both of the NAT tails. We emphasize that within their active jet run periods, N2, N3 and N4 are identical at a given time, $t$, so long as the jets remain active. The start times and jet deactivation times differ according to the criteria just mentioned. These run extensions provide a control group to evaluate the explicit consequences of the shock encounters.

\begin{table}[ht]
\begin{center}
\begin{tabular}{| c | c | c | c | c | c |}
\hline
Name & Restart time & AGN off time & Shock Strength & Shock Speed & Shock \\
	& (Myr) & (Myr) & (Mach \#) & (kpc/Myr) & Compression \\ \hline
M2 & 409.9 & 518.1 & 2   & 2.045 & 2.286 \\ \hline
M3 & 508.3 & 546.6 & 3   & 3.068 & 3.000 \\ \hline
M4 & 557.5 & 586.0 & 4   & 4.091 & 3.368 \\ \hline
\hline
N2 & 409.9 & 518.1 & N/A & N/A   & N/A   \\ \hline
N3 & 508.3 & 546.6 & N/A & N/A   & N/A   \\ \hline
N4 & 557.5 & 586.0 & N/A & N/A   & N/A   \\ \hline
\end{tabular}
\end{center}
	\caption{Simulation parameters for different ``extensions'' or ``runs'' in this work. Ages are measured from AGN source activation, with ``restart time" referring to the age of the AGN when each specific simulation extension was started. Shock attributes are representative of the plane shock in the background ICM.}
	\label{tab:param}
\end{table}

In the simulations restarted to launch shocks we fixed ICM flow conditions along the $-\hat{z}$ boundary to represent the post shock state of a plane shock with Mach number, $\mathcal{M} =$ 2, 3 or 4, propagating into the undisturbed ICM in the $+\hat{z}$ direction. The restart times for run extensions M2, M3 and M4 were chosen so that the total age of the NAT was approximately the same at the end of each  full shock encounter. The previously outlined inflow boundary conditions on the $-\hat{x}$ boundary were maintained, so that the ICM wind continued to flow in from that boundary. 
The ICM density, pressure and $v_z$ velocity component were fixed in the $-\hat{z}$ boundary using standard shock jump conditions. Explicitly, using subscript ``b'' to represent boundary values, $\rho_b$, $P_b$ and $v_{z,b}$ became for each Mach number, $\mathcal{M}$,
\begin{align} 
\label{eq:RH1} \rho_b & = \frac{4 \mathcal{M}^2}{\mathcal{M}^2 + 3} ~ \rho_w, \\
\label{eq:RH2} P_b & = \frac{5 \mathcal{M}^2 - 1}{4} ~ P_w, \\
\label{eq:RH3} v_{z,b} & = \frac{3}{4} \frac{\mathcal{M}^2 - 1}{\mathcal{M}}\sqrt{\frac{5 P_w}{3\rho_w}}.
\end{align}
The tangential velocity components, $v_x$ and $v_y$ were continued from inside the domain, so across the shock. In each of the  M2, M3 and M4 run extensions, the AGN jets were switched off when the ICM shock made first contact with the nearer of the two tails (corresponding to the jet pointed in the $-\hat{z}$ direction). Those times are also listed in Table \ref{tab:param}. The shock speed in Table \ref{tab:param} is $v_{z,b}$ from Eqn. \ref{eq:RH3}, while the compression figure in Table \ref{tab:param} is the ratio, $\rho_b/\rho_w$ derived from Eqn. \ref{eq:RH1}.

\section{NAT-Shock Interaction Dynamics} \label{sec:DynShk}

Our merger shocks were initialized with speeds ($v_{s,w}$) of 2, 3 and 4 times the sound speed of the undisturbed background ICM wind, $c_w = 1000$ km/s, but the specifics of the shock interaction within the NAT obviously depended on the local state of the tail plasma being shocked. The density structure within the tails is especially important during propagation through the NAT, as we lay out quantitatively below. We outlined in \S \ref{sec:NAT} how the pre-shock tail density structures are made highly heterogeneous by the unsteady propagation of the AGN jets penetrating into the tails and the associated deposition of jet plasma into the tails. Although the tails maintain approximate pressure balance with the surrounding ICM, the fact that they include hot, lower density jet plasma leads them to be generally less dense than their surroundings. But, that density is not at all uniform. 

To clarify this behavior it is convenient to examine the tails in the \textbf{N3} simulation at $t \approx 541$ Myr. Figure \ref{fig:BnC165} shows volume renderings of the mass fraction, $C_j$ and magnetic field strength at this time, while figure \ref{fig:N3150} illustrates the resulting synthetic synchrotron observations. Defining the NAT as the volume within which $C_j > 0.01$ we find a median density $\sim0.31 \rho_w$, with about half the volume having densities between $0.1\rho_w$ and $0.5\rho_w$. At the same time, because of the approximate pressure equilibrium, the lower density regions tend to be hotter, so have larger sound speeds, $c$. Accordingly, we find in this case, 50\% of the NAT volume had $c \geq 2c_w$, 11.6\% had $c \geq 3c_w$ while 4\% had $c \geq 4c_w$. As we work out in the following subsection, the shock generally propagates faster in lower density volumes, but, because of the higher sound speed, propagates at reduced Mach number. 

So, in general terms the shocks will propagate faster through the tails than around the tails, but shock compression within the tails will generally be less. Just as important, however, the fact that the density structure within the tails is highly heterogeneous, means that as it passes through a tail the shock surface is deformed. In addition, as we also outline below, the density variations lead the shock to generate substantial shear (vorticity) within the tail. On top of the consequences of the heterogeneous density structure are the influences from pre-shock turbulent motions in the tails. This adds also to the vorticity generation. We examine these consequences below.

We comment, in passing, on the presence of a low density bridge evident between, but near the ends of the tails in figure \ref{fig:BnC165}. As discussed in \cite{on19a} this feature is a remnant of the early evolution of the NAT, when the jets were first deflected by the wind. As a dynamical feature this structure is quite long lived. But, since it became detached from the jets long ago, decay of magnetic fields as turbulence decayed and radiative cooling of CRe in the bridge made it virtually undetectable as a synchrotron source, by this time (cf. figure \ref{fig:N3150}). Shock propagation is accelerated through this bridge in comparison to other, denser regions between the tails.

\subsection{Shock Strengths in the Tails and Shock Induced Vorticity} \label{sec:ShkVort}

As just outlined, complex structure within the simulated NAT prior to its encounter with the initially plane shock leads the shock within the NAT to become complex and to generate significant dynamical complexity. Broadly speaking, the shock propagates faster, but is weaker within the tails than in the undisturbed medium. The analytic analysis of \cite{pj11} provides a simple means to explore this behavior quantitatively, if only approximately in this setting. Specifically, \cite{pj11} used a 1D Riemann analysis to calculate how a shock penetrating through a low density cavity crushes the cavity and transforms  it into a toroidal vortex ring, analogous to a smoke ring. This behavior had been studied in laboratory fluids \citep[\eg][and references therein]{ranjan08} and also was suggested in the context of shocked fossil RG lobes \citep{enss02} as candidates to provide CRe to form radio relics. \cite{nolt19,nolt19b} demonstrated this behavior explicitly in MHD simulations of shock impact on lobed RGs. 

Specifically, when a shock with velocity, $v_{s,i} = \mathcal{M}_i c_i$, travelling through a stationary medium  with density $\rho_i$, enters a low density a cavity ($\rho = \rho_c < \rho_i$), initially in pressure balance with the unshocked  external medium, the shock propagates into the cavity with velocity $v_{s,c} = M_c~ c_c > v_{s,i}$, but $\mathcal{M}_c < \mathcal{M}_i$, while the initial cavity-ICM contact discontinuity (CD), follows at speed, $v_\text{CD} < v_{s,c}$. A rarefaction also propagates backwards into the post-shock external medium, although this has fewer consequences. Given a cavity density ratio $\delta = \rho_c/\rho_i < 1$, \cite{pj11} showed that Riemann invariants and standard shock jump conditions can be utilized to derive the following equation relating the initial Mach number $\mathcal{M}_i$ to the Mach number in the cavity, $\mathcal{M}_c = \mu \mathcal{M}i$, as (with $\gamma = 5/3$):
\begin{equation}\label{eq:cavity}
1 = \frac{\mu^2\mathcal{M}^2_i - 1}{\mu \delta^{1/2}\left(\mathcal{M}^2_i -1\right)} - \frac{\left(5\mathcal{M}^2_i - 1\right)^{1/2} \left(\mathcal{M}^2_i + 3\right)^{1/2}}{\mathcal{M}^2_i - 1}  \left[1 -\left(\frac{5\mu^2\mathcal{M}^2_i - 1}{5\mathcal{M}^2_i - 1} \right)^{1/5}\right].
\end{equation}

\begin{figure}[ht]
\centering
\includegraphics[width=\textwidth]{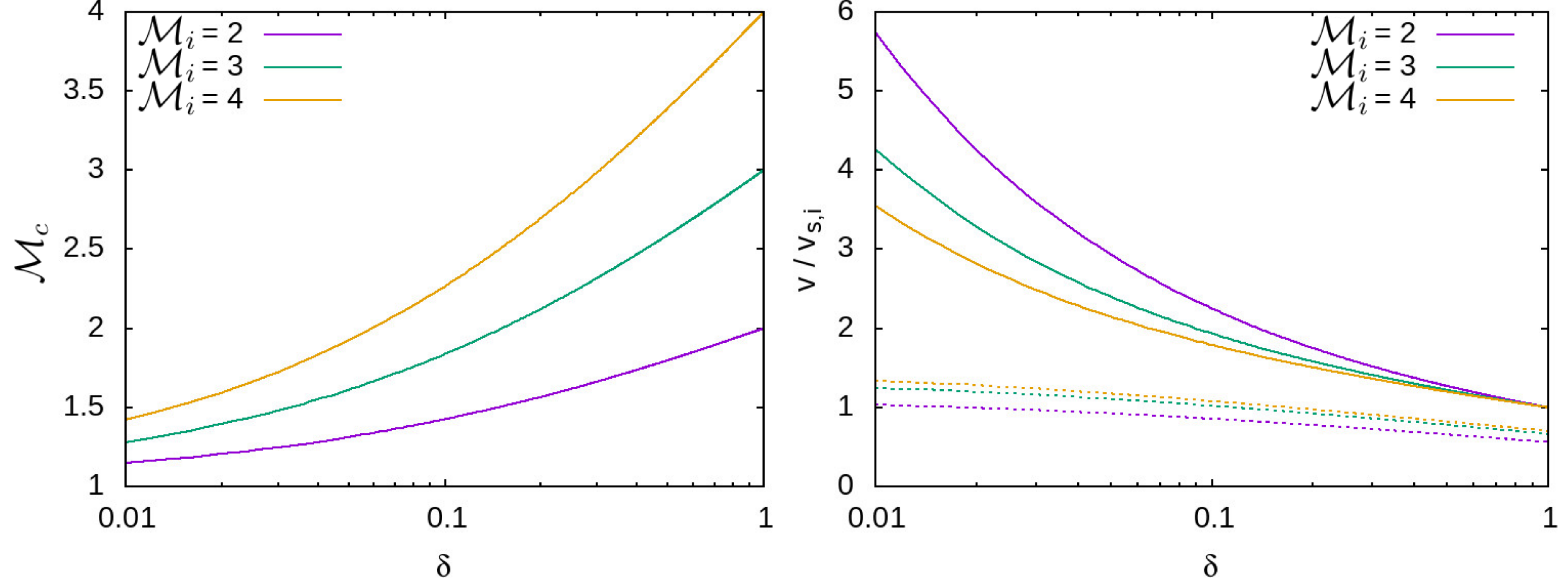}
\caption{Properties of the external shock strengths we use as they propagate within a low density cavity as a function of the density ratio $\delta  = \rho_b/\rho_i$. On the left: shock strength within the cavity $M_c$. On the right: wave speeds in the cavity relative to ICM shock speeds. Solid lines are $v_{s,c}/v_{s,i}$, while the dotted lines are $v_\text{CD}/v_{s,i}$.}
\label{fig:bubble}
\end{figure}
On the left side of figure~\ref{fig:bubble}, we illustrate numerically obtained solutions to this equation for $\mathcal{M}_c$ as a function of the density ratio, $\delta$, for $\mathcal{M}_i=$ 2, 3, 4. By definition, the speed of the shock in the low density medium, $v_{s,c} =v_{s,i} ~ \mu / \sqrt{\delta}$. The speed of the CD, $v_\text{CD}$, equals the speed of the post-shock flow velocity; namely,
\begin{equation}
v_\text{CD} = \frac{3(\mu^2 \mathcal{M}_i^2-1)}{4\sqrt{\delta} \mu \mathcal{M}_i^2} v_{s,i}.
\end{equation}
We plot the ratios, $v_{s,c}/v_{s,i}$ and $v_\text{CD}/v_{s,i}$ on the right side of figure~\ref{fig:bubble}. In all cases with $\delta <1$, the propagation time of the shock is shorter through the low density medium than in the background, and the shock propagates fastest through regions with lowest density. As the density in the cavity decreases, the strength of the shock asymptotically approaches $\mathcal{M}_c = 1$, although the actual speed of the shock increases, because the local sound speed is greater by a larger amount. 

\cite{pj11} simulated a shock impacting a spherical cavity of radius, $r_c$, showing how an approaching plane shock encounters first the tangent point of the cavity, and then because $v_{s,c} > v_{s,i}$, passes through the cavity center faster than around the periphery. The CD follows the cavity shock at a rate, $v_\text{CD} \sim v_{s,i}$. The resulting shear along the cavity boundary transforms the spherical cavity into a stable, toroidal vortex in a  time of order $2 r_c/v_\text{CD}$. As evident in figure~\ref{fig:bubble} for our situation, the CDs associated with the impacting shocks all travel at speeds on the order of the incident shock speed. In particular, $v_\text{CD} \approx 1.34v_{s,i}$ for $\mathcal{M}_i = 4$ and $\delta = 0.01$, or $v_\text{CD} \approx 0.5 v_{s,i}$ for $\mathcal{M}_i = 2$ as $\delta \rightarrow 1$.

One important consequence not yet emphasized of the heterogeneity of the shock propagation through the first tail comes from the fact that the shock surface becomes strongly curved. This, on average, leads to a divergent post-shock flow, which locally weakens the exiting shock compared to the entering shock. We find in our simulations that the exiting Mach numbers across most of the shock front are $\sim5$-10\% lower than the Mach number of each shock before it encountered the near tail. Shock sections that are extremely distorted can have the Mach number at impact on the second tail reduced by as much as ~$\sim50$\%. Therefore, the strength of the shock upon encountering the second tail is everywhere weaker than it was before interacting with the first tail.

\begin{figure}[ht]
\centering
\includegraphics[width=\textwidth]{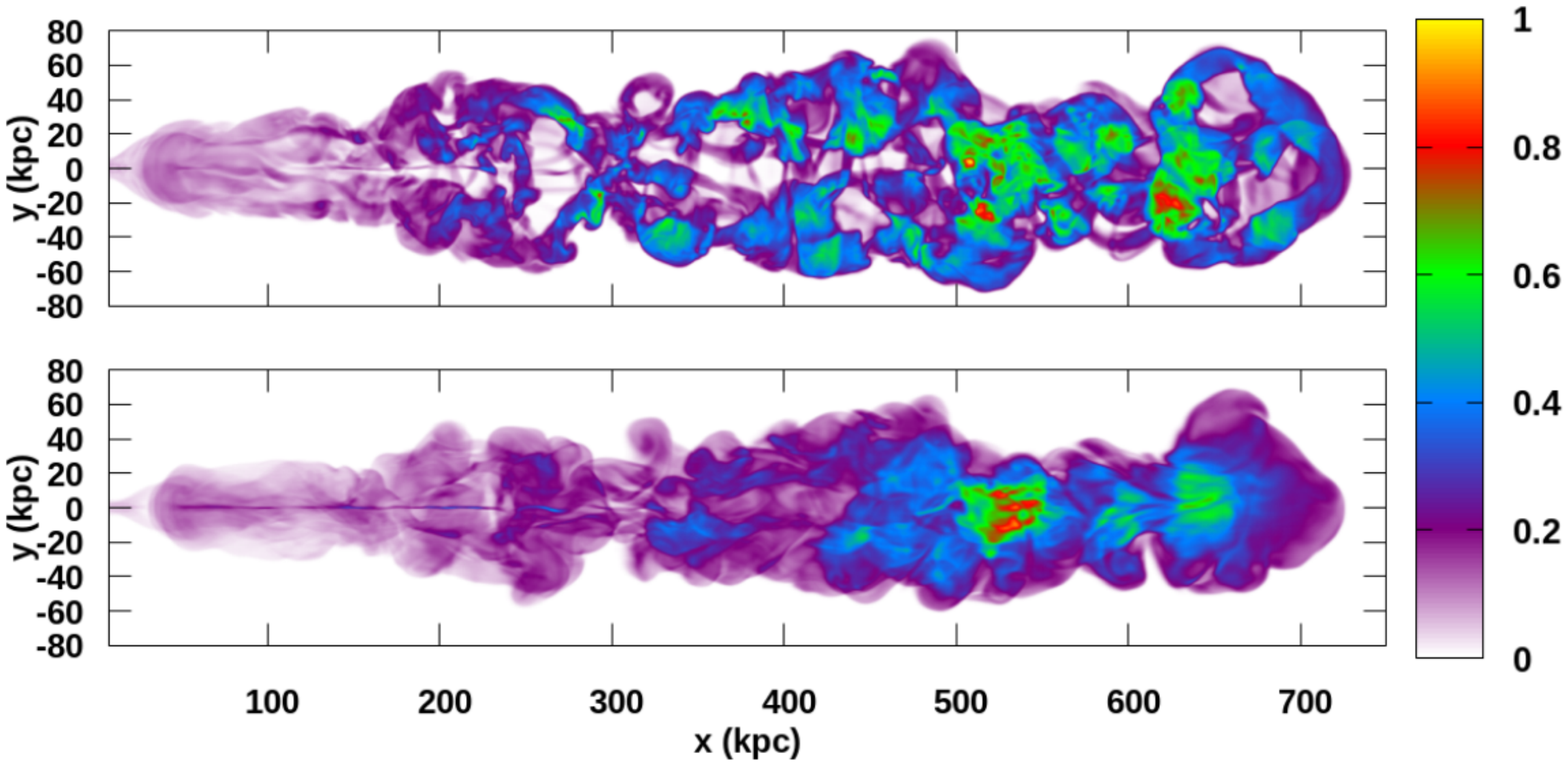}
\caption{Jet mass projected along the $\hat{z}$ axis through the first-encountered ($-z$)~tail at $t \approx 603$~Myr, about 30~Myr after the Mach 3 shock  has completed passage through the tail in the \textbf{M3} run. Top: From run \textbf{M3} (includes the shock). Bottom: run \textbf{N3} (no shock). The initial shock plane was parallel to the image plane. The projected mass in each image is scaled to the maximum value of both images.} \label{fig:JetMass}
\end{figure}
To explore the shock generation of vorticity within the tails it is useful to consider them as a series of spheroidal plumes (cavities) sequentially inflated by the jets. According to the analysis associated with Eqn. \ref{eq:cavity}, shock passage transforms each of these plumes into a vortex ring, with smaller scale holes developing through lower density volumes where the shock and CD propagate significantly faster. These effects are evident in figure~\ref{fig:JetMass}, which displays the projected jet mass\footnote{The product of $C_j$ and $\rho$ integrated along a line of sight} through one of our tails at the same point in time in simulations \textbf{M3} and \textbf{N3}, some $\sim 30$~Myr {\it after} the shock has completed passage through the tail. It is apparent that through much of the tail, jet plasma has been swept towards the perimeters by vortical motions, particularly from volumes that were initially at lower density. For example, in the bottom image (from simulation \textbf{N3}, so absent a shock interaction) the right most plume of the tail has denser plasma towards its left (centered around $x \approx 640$~kpc) relative to the rest of the plume. Thus, the upper image from simulation \textbf{M3} is explained if the shock had traveled relatively slower through this region. As a result, much of the shocked plume has bifurcated into separate rings surrounding this denser plasma.

As noted above, the  NAT tails were very turbulent as a consequence of their formation. The shock behaviors just discussed enhance this turbulence. Shock enhancement of turbulence is a well-known behavior, of course \citep[\eg][]{giac07,ji16}. The smaller scale ring-like structures visible in figure \ref{fig:JetMass} end up being relatively short-lived features that are torn apart by turbulent motions. The larger scale vortical motions are longer lived, and in the long term they replace nearly all the jet plasma near the tail midplane with ``diffuse plasma'' characterized by $C_j \ll 1$. Thus, a projected jet mass image like the upper panel in figure~\ref{fig:JetMass} evolves towards something resembling an elongated ring outlining the original distribution (i.e., tracing the bounds of the distribution in the lower panel of figure~\ref{fig:JetMass}). As it turns out in this simulation, the transformation time towards this morphology is coincidentally roughly the cooling time for radio luminous reaccelerated CRe ($\sim100$~Myr, see~\S~\ref{sec:SynObs}), so it may be unlikely that a shocked tail with subsequently  relaxed post-shock vortices would be observable as a radio source. Also, although the density configuration appears coherent at this time when viewed along the direction of shock propagation, by the time the shock-generated vortices relax, the tail plasma is no longer co-planar. Immediately after shock passage, the transverse extent of the tail is compressed compared to the pre-shock state, but the turbulence mentioned above gradually disperses tail plasma along this direction into detached filaments that appear uncorrelated from other viewpoints.

\subsection{Magnetic Field Strength and Topology Pre and Post Shock} \label{sec:ShkMag}

Before examining the magnetic field  distributions in our simulated NAT, both before and after the shock encounters, it is useful to  recall that our ``pristine'' ICM ($C_j = 0$) was, for simplicity purposes, unmagnetized, and that the peak magnetic field strength in the jets as they emerged from their launch cylinder was $1~\muG$. Accordingly, using the launched jet field as a standard for comparison, we somewhat arbitrarily divide at $B  = 1 \muG$ ``weak'' from ``strong'' fields. By itself, this designation does not establish the dynamical impact of the field, although in these simulated environments field strengths below $1 \muG$ are unlikely to have significant dynamical impact.

Given the highly turbulent character of the NAT tails, it is no surprise that the magnetic fields in those tails were largely disordered before arrival of the ICM shock. That character was verified explicitly in \cite{on19a} and is evident in the volume rendering of the magnetic field  from the \textbf{N3} run at $t \approx 541$ Myr shown in figure \ref{fig:BnC165}. Indeed, averaged over the entire NAT volume, the magnetic field in the tails at that time is essentially isotropic. On the other hand the radio synchrotron image in figure \ref{fig:N3150} clearly reveals magnetic filaments that tend to align with the tails. Comparison of figures \ref{fig:BnC165} and \ref{fig:N3150} suggests that the filamentary synchrotron features relate especially to relatively stronger magnetic field regions that, as \cite{on19a} pointed out, result from episodic jet disruption events within the tails. Those dynamics maintained the existence of magnetic field structures stretched out roughly along the tails over a considerable tail length. Indeed, analysis of magnetic fields stronger than 1 $\muG$, which corresponds roughly to the brighter synchrotron emitting regions in figure \ref{fig:N3150}, shows a strong directional bias along $\hat{x}$; that is, alignment with the tails, which in our simulated scenarios biases their alignment towards the plane of incident shocks. The corresponding bias in polarization of radio synchrotron emission would be orthogonal to the wind direction (although with substantial scatter). Farther downwind, turbulence prevails, isotropizing the field structures. Even farther down the tails, turbulence  and magnetic fields both relaxed into disordered but weaker structures. These magnetic field evolution behaviors were also highlighted by \cite{on19a}, in their analysis of the radio polarization properties of the pre shock NAT under discussion. They pointed out in that context, in particular, the contributions of stronger, filamented fields, but also the increasingly patchy, less distinctive character of polarization farther down the tails.

As discussed in \S \ref{sec:ShkVort}, the dynamic consequences of shock passage through the NAT can be expressed in terms of compression and added shear (vorticity, turbulence). Both impact the magnetic field strength and topology in the NAT. As is well known, compression enhances the component of the magnetic field perpendicular to the local direction of compression (shock propagation in this case), but not the aligned component. So, even a random magnetic field distribution, after being shocked, will be preferentially biased towards the shock plane. For our shock-NAT interactions we chose a particularly simple scenario in which an incident plane shock plane was roughly tangent to the pre shock tails. That was expected to lead to the simplest post shock structures with maximum alignment of the magnetic field along the plane of the incident shock. Indeed, as just discussed, the stronger pre shock magnetic field structures were already biased towards tail alignment. Consequently, those stronger field structures tended to be more nearly aligned to the incident shock plane than the more nearly isotropic weaker magnetic fields. Shock compression preferentially strengthened the initially stronger fields, and enhanced the tendency of the post-shock magnetic fields to lie in the plane of the incident shock, as expected. 

{\it{On the other hand}}, shock compression is a single, abrupt transition, while, because of the heterogeneous and turbulent character of the NAT before shock impact, shock passage in our numerical experiments greatly enhanced turbulence within the tails. Consequently, field line stretching (in effect the ``turbulent dynamo'') continued to strengthen the magnetic fields for 10's of Myr after shock passage \citep[see also, \eg][for other analogous contexts]{ji16}. The post shock turbulence was neither uniform nor steady, nor was it independent of the strength of the incident shock. Detailed analysis is beyond the scope of this study, but we can {\underline{very}} roughly characterize post shock eddy scales as $\sim \ell_b$ and velocities as  $\sim (1/2)~ c_w$. Magnetic field amplification peaked in the first-encountered, $-z$~tail about 25, 20 and 13~Myr after the shock had completed passage through the tail for the \textbf{M2}, \textbf{M3} and \textbf{M4} simulations we carried out. These times correspond, as well, to the maximum radio fluxes in our synthetic observations of the shocked NAT (see \S~\ref{sec:SynObs}). In addition they roughly match when the post shock CD mentioned in \S \ref{sec:ShkVort} completely traversed the tail and post-shock vortex structures (see \S~\ref{sec:ShkVort}). Subsequent to those times, both the shock generated turbulence and amplified magnetic fields relaxed and diffused outwards, so that over timescales similar to those associated with the amplification the magnetic field distributions roughly approached those of the pre shocked tails.

We also make the point here that after running through the first tail and generating the behaviors just outlined, each shock was significantly weakened and very distorted. The subsequent interaction with the second, $+z$~tail was, consequently, weaker and not nearly as clean. Not only compression, but also the generated turbulent dynamos were less effective in the second, $+z$, than in the initial, $-z$, tail encounter. We shall see in the following section, that the synchrotron emission responses to the shock encounter with the second tail are then much weaker, as well.

\begin{figure}[ht]
\centering
\includegraphics[width=\textwidth]{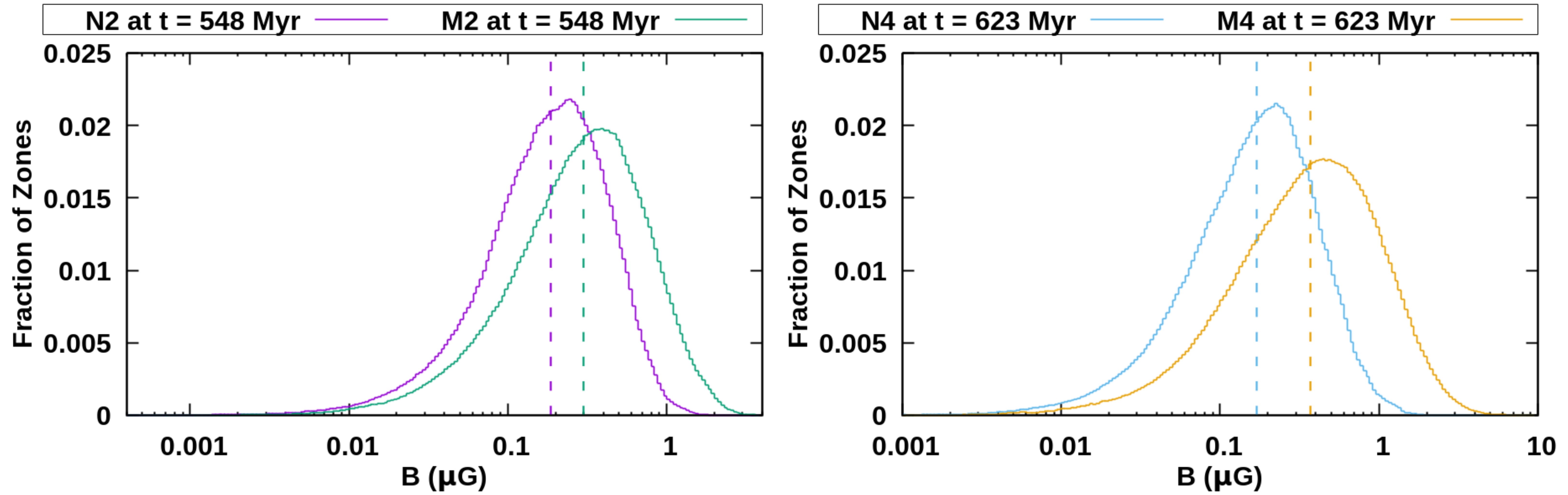}
\caption{Probability distributions of magnetic field strength for zones (cells) in the $-z$~tail with $C_j \geq 0.01$  at fiducial times described in the text and shown above the plots for shocked and unshocked cases. Zones are sorted into logarithmically spaced bins with 50 bins per dex. The dashed vertical lines show the median for each distribution. Left: $\mathcal{M} = 2$; Right: $\mathcal{M} = 4$.}
\label{fig:Bdist}
\end{figure}

A common and useful measure of magnetic field properties is the spatial probability distribution, or ``PDF''. Figure \ref{fig:Bdist} illustrates the PDF of magnetic field intensities in the first-encounter, $-z$ tail for the $\mathcal{M} = 2$ shock (\textbf{M2} run) (left panel) and the $\mathcal{M} = 4$ shock (\textbf{M4} run) at the times when their magnetic fields are most strongly amplified by post shock turbulence ($t \approx 548$ Myr and $t \approx 623$ Myr, respectively). For comparison, the figure also includes the magnetic field distributions at those same times for respective NAT formation extensions, \textbf{N2} and \textbf{N4}, in which there are no shocks. The distributions are approximately log-normal. On the left we see that the median magnetic field strength is enhanced in response to the  $\mathcal{M} = 2$ shock encounter by a factor $\sim 1.6$, to something close about $0.3\muG$,  while the median field strength is enhanced following the $\mathcal{M} = 4$ shock by a factor $\sim 2$ to about $0.4\muG$. In both cases the PDFs are also broadened, but by somewhat smaller factors easily estimated from the figure.

\section{Analysis of Synthetic Observations of the Shocked NAT} \label{sec:SynObs}

We used data from the simulations in Table \ref{tab:param} to compute and analyze synthetic synchrotron radio images. Spectral intensity, $I_\nu$, and spectral index, $\alpha_{\nu1,\nu2}$, maps were made at redshifted reference frequencies\footnote{The frequencies are a factor $1+z = 1.2$ higher in the source rest frame.}, 150 MHz, 325 MHz, 600 MHz, 950 MHz and 1.4 GHz following the same methods as in \cite{on19a}. We examined intensity and spectral index ($\alpha$) maps at two time snapshots from simulations \textbf{M2}, \textbf{M3} and \textbf{M4} (after the shock has passed through the first-encounter, $-z$~tail and after passing through both tails). Resulting intensity images at 150 MHz and 950 MHz along with $\alpha_{150,325}$ and $\alpha_{950,1400}$ maps  viewed down the $+\hat{y}$ direction are shown in  figures~\ref{fig:M2obs1}, \ref{fig:M2obs2}, \ref{fig:M3obs1}, \ref{fig:M3obs2}, \ref{fig:M4obs1} and \ref{fig:M4obs2}. We also used synthetic images from throughout the \textbf{M2}, \textbf{M3} and \textbf{M4} simulations, as well as complementary images from \textbf{N2}, \textbf{N3} and \textbf{N4}, to compute the time evolution of integrated radio fluxes and spectra shown in figures~\ref{fig:M2SpcEvo}, \ref{fig:M3SpcEvo} and \ref{fig:M4SpcEvo}.

In $\sim\mu$G scale fields found in our simulated NAT, synchrotron radiation at radio frequencies $\nu \sim 0.1$-$1$~GHz comes primarily from CRe with Lorentz factors, $\Gamma_e\approx p/(m_e c)~ \sim 10^4$~--~$3\times10^4$. The lifetimes, $\tau_{rad}$, against synchrotron and iC emissions can be expressed \citep[\eg][]{szin99}, 
\begin{equation}
    \tau_\text{rad}\approx 108 \left(\frac{10^4}{\Gamma_e}\right) \left(\frac{4.67 \muG}{B_\text{eff}}\right)^2~\text{Myr},
\end{equation}
where $B_\text{eff}^2 = B^2 + B_\text{CMB}^2$ with $B_\text{CMB} \approx (1+z)^2~3.24\muG = 4.67\muG$ as the equivalent strength of the CMB for our assumed redshift, $z=0.2$. In a $1\muG$ magnetic field with $B_\text{eff} \approx 4.67\muG$ at $z = 0.2$, iC is the dominant radiative cooling mechanism for CRe. After shock compression there are volumes where the field is about $5\muG$, so where iC and synchrotron cooling rates are roughly equal and the CRe lifetimes are $\sim 50~(10^4/\Gamma_e)$ Myr. Fields approaching $10\muG$ are attained in some regions in the \textbf{M4} simulation. So, the nominal radiative lifetimes can be as short as $\sim 20~(10^4/\Gamma_e)$ Myr in these spots, although the duration of such strong field regions is similarly short (see \S~\ref{sec:ShkMag}).

We found in \citet{on19a} that so long as the AGN source remained active, the radio appearance, integrated radio luminosity and spectrum of the unshocked NAT became roughly steady state after $\sim 200$ Myr, by which time the NAT morphology was fully developed and CRe injected from the AGN during the early and transitional formation stages had ``aged'' below detectability at our reference frequencies. For the subsequent duration of the jets' activity, the NAT largely resembled figure~10 in that work. The ``visible'' extent of the tails (in the $x$-dimension) stayed $\sim400$~kpc at 150~MHz and $\sim300$~kpc at 950~MHz. Older tail plasma downstream still contained relativistic CRe, but, in the absence of any Fermi II reacceleration, those CRe had cooled below energies capable of radiating at our reference radio frequencies. In the present context shock fronts pass through the NAT, activating this plasma once again, to radiate in the radio band. That ``rebirth'' results both from compression and DSA reacceleration of CRe to higher energies and also from magnetic field amplification discussed in the previous section. The latter effect increases the synchrotron critical frequency, $\nu_c\propto \Gamma_e^2 B_\perp$, of CRe previously not energetic enough to be seen at radio frequencies.

\subsection{Synthetic Observations from the Mach 2 Shock Encounter} \label{sec:M2obs}

\begin{figure}[ht]
\centering
\includegraphics[width=\textwidth]{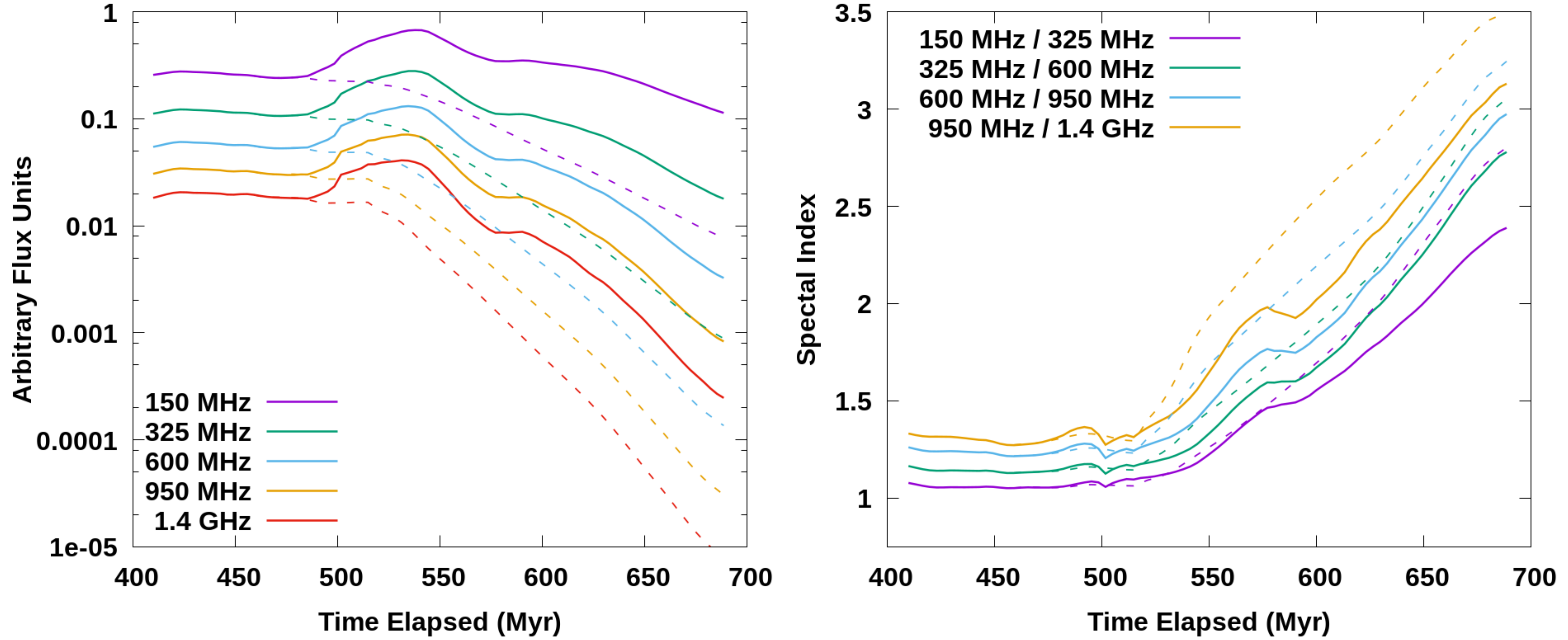}
\caption{Evolution of integrated radio emission in the \textbf{M2} (shocked) and \textbf{N2} (unshocked) runs. Solid lines are from \textbf{M2} and dashed lines are from \textbf{N2}. Left: Time evolution of integrated fluxes (arbitrary units) at 150~MHz, 325~MHz, 600~MHz, 950~MHz and 1.4~GHz. Fluxes are sampled approximately every 3.3~Myr, starting at $t\approx 410$~Myr. Right: Time evolution of integrated spectral indices, $\alpha_{\nu 1,\nu 2}$, between adjacent frequencies on the left, sampled at the same time interval. The AGN jet source turned off at $t \approx 518$ Myr} \label{fig:M2SpcEvo}
\end{figure}
We begin our analysis of radio emissions from the shocked NAT with the \textbf{M2} simulation, which has the weakest and thus slowest moving shock. Because the shock is weak, reacceleration of CRe (mostly by adiabatic compression in this case) barely compensates for radiative cooling during the shock crossing. It also turns out that this shock exhibits the largest relative variations in shock speed within the tails in response to density variations (see \S \ref{sec:ShkVort}). As one consequence,  where densities are low in the $-z$ tail the \textbf{M2} shock  gets more obviously ahead of the associated shock traversing the denser surroundings than is the case for the stronger, \textbf{M3} or \textbf{M4} shocks. 

The \textbf{M2} shock first encounters significant amounts of tail plasma around $t \approx 480$~Myr and finishes its pass through  the first, $-z$ tail, about 40~Myr later, shortly after the jet turns off at $t\approx 518$~Myr. Once the shock has passed through the $-z$~tail, radio emissions in that tail continue to brighten for another $\sim 25$~Myr. This is mostly in response to turbulent magnetic field amplification (see \S~\ref{sec:ShkMag}), although compression by the shock of fresh jet plasma released from the AGN just prior to shutting off (as the shock passage is underway) also enhances the emissions. Even while fluxes increase during this time, spectral indices begin to steepen  noticeably as recently-shock-reaccelerated CRe resume cooling. The peak integrated fluxes at the five frequencies plotted in figure~\ref{fig:M2SpcEvo} are not exactly simultaneous, but still occur within 7~Myr of each other. They peak first at 1.4~GHz and then at sequentially lower frequencies. Across the spectrum, the peak integrated fluxes in \textbf{M2} are almost a factor of 4 greater than at the comparable times in the paired, unshocked NAT extension, \textbf{N2}. The time spectral evolution (spectral index, $\alpha$) on the right side of figure~\ref{fig:M2SpcEvo} shows that the integrated spectrum mostly maintains the same curvature throughout the shock encounter with the $-z$ tail.

\begin{figure}[ht]
\centering
\includegraphics[width=\textwidth]{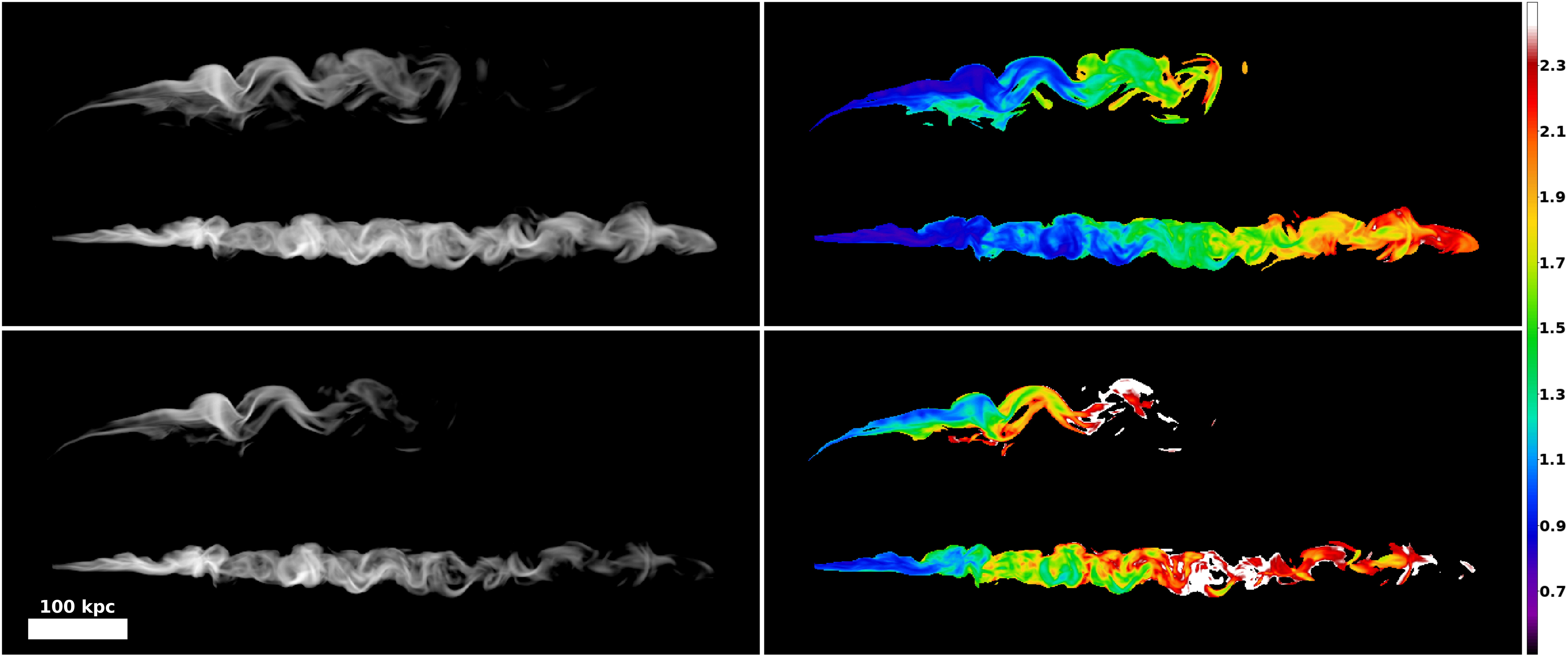}
\caption{Radio images from \textbf{M2} at $t \approx 525$~Myr viewed orthogonal to the plane containing the AGN jets and the ICM shock normal. Top Left: $I_{\nu,150}$. Bottom left: $I_{\nu,950}$. Top right: $\alpha_{150,325}$. Bottom right: $\alpha_{950,1400}$. Intensity images are in arbitrary units on a log scale, with a dynamical range of 500:1. The $\alpha$~maps are generated using only pixels having $I_{\nu}$ within a factor 500 of the brightest pixel at the higher frequency. At this time the shock lies between the two tails.} \label{fig:M2obs1}
\end{figure}
Figure~\ref{fig:M2obs1} shows intensity and $\alpha$~maps shortly after the shock has passed through the $-z$~tail plus a segment of recently released jet plasma initially just above that tail, but at this time almost merged into it. The highly filamentary intensity images in the figure are dominated by two bright spots at both frequencies shown. The first spot (to the left) consists of the just-mentioned remains of the now terminated jet and also what \cite{on19a} called the disruption zone in the $-z$ tail, where slightly older, but still coherently focused jet plasma  was in the midst of a disruption event when the shock came through. The other bright spot is a collection of filaments about 90~kpc down the tail from the first spot, containing  magnetic fields enhanced during a previous major disruption event. Both of these bright spots can be identified with clumps of strong magnetic filaments in the unshocked tails on the right side of figure~\ref{fig:BnC165}. 

A spectral gradient along the tails tracing the age of jet plasma is evident from the $\alpha$ maps in the figure. The absence of spectral flattening across the shock, on the other hand, suggests that adiabatic compression is the dominant reacceleration mechanism rather DSA. In fact, comparing $\alpha$ in the shocked tail to the still-visible portion of the unshocked tail, there is, indeed, a slight $\alpha$ offset consistent with compression adiabatically shifting a CRe distributions with convex spectral form to higher energies. Similarly, individual brighter filaments stand out with slightly flatter spectra relative to adjacent emission as a result of lower energy CRe slightly boosted by compression and radiating at a given frequency in comparatively stronger magnetic fields.  

However, in the high frequency $\alpha$~map, the spectrum farthest down the tail (on the right) is actually slightly flatter than sections just to the left, suggesting some role for DSA, even though the intensity, $I_\nu$, is relatively weak, because the CRe population is relatively low.
This spectral flattening towards the tail end coincides with regions there with relatively high mass density, so predominantly ICM. As outlined in \S \ref{sec:ShkVort}, the shock slows down in denser regions, but the shock Mach number is enhanced. Emissions from portions of this region have near power law shape with $\alpha \sim 1.7$ across our reference frequency domain. This is consistent with DSA reacceleration associated with a shock Mach number, $\mathcal{M} \sim 1.6$ in this region. Thus, the shock penetrating this tail region is only moderately weaker than the incident shock in the wind. 

\begin{figure}[ht]
\centering
\includegraphics[width=\textwidth]{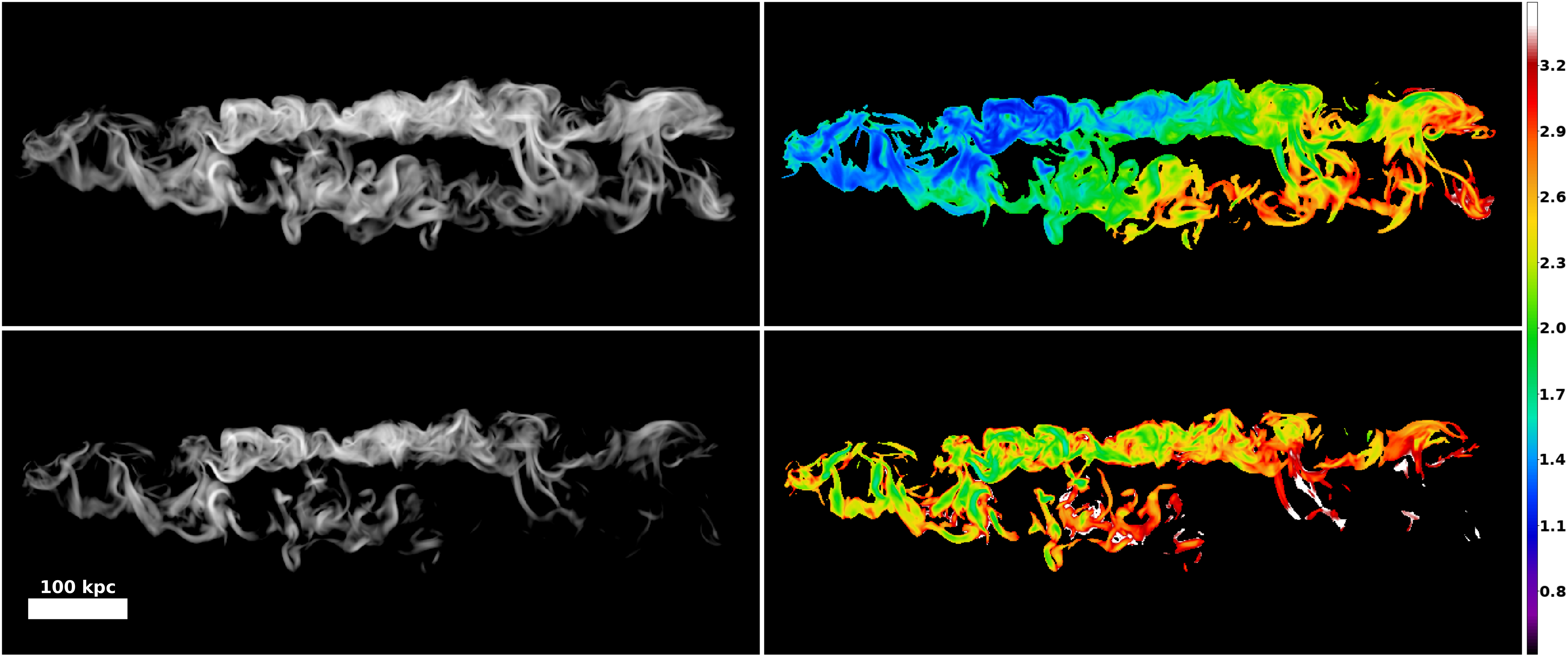}
\caption{Same as figure~\ref{fig:M2obs1} for \textbf{M2}, but at $t \approx 617$~Myr, when the radio emissions peak and just after the shock completes both tail passages. Note the different color scale for the $\alpha$~maps compared to figure \ref{fig:M2obs1} on account of more advanced cooling of the CRe at this later time.} \label{fig:M2obs2}
\end{figure}

After the $-z$ tail reached peak brightness in response to shock compression and related turbulent magnetic field amplification, and while the shock propagated between the two tails, radiative cooling dominated the CRe and synchrotron emission evolution in the $-z$ tail. At the highest frequencies (600~MHz and above) integrated fluxes faded below their pre-shock levels by $t \approx 565$~Myr. During this time interval the CRe in the $+z$ tail continued to cool radiatively, so that by the time the shock encountered the second tail, synchrotron emissions were no longer visible there for $\nu \ge 600$ MHz. As the shock began encountering significant amounts of plasma in the $+z$~tail,  around $t\sim 575$~Myr, the combined fluxes from both tails stabilized and remained roughly constant. The new particle acceleration at the shock in the $+z$ tail approximately balanced radiative losses within the $-z$ tail and the still unshocked portions of the $+z$ tail. This pause in the flux evolution ended once the shock had run through the bulk of the plasma in the $+z$ tail, so that except for weak turbulent magnetic field amplification in the $+z$ tail, decay processes prevailed after the shock exited the second tail.

The shock front entering the $+z$ tail was somewhat weakened and significantly distorted after its interaction with the $-z$ tail. As a result, the $+z$ tail was even less cleanly modified by the shock compared to the $-z$~tail. On the other hand the shock Mach number was consistently around $\mathcal{M} \sim 1.5$ throughout much of the $+z$ tail. For CRe accelerated by DSA that corresponds to a synchrotron spectral index, $\alpha = 2.1$. Just prior to being shocked, higher energy CRe in the $+z$~tail had cooled enough that synchrotron emissions were no longer visible at the higher frequencies we considered.  Post shock, the high frequency emissions visible in the $+z$ tail in figure~\ref{fig:M2obs2} are largely consistent with DSA from a shock of Mach number, $\mathcal{M}\sim 1.5$. Because the lower frequency pre shock spectra from the $+z$ tail were mostly flatter spectra  than this, but steepened towards the end of the tail, the post shock $+z$ synchrotron spectral distribution included a spatial spectral gradient pattern preserving the signature of pre shock radiative aging. 

Also because the shock encountering the $+z$ tail was weakened by its passage through the $-z$ tail, the high frequency emissions in the $+z$ tail had steeper spectra. In the roughly $100$ Myr that transpired between between shock exit from the $-z$ tail until shock exit from the $+z$ tail CRe in the $-z$ tail aged substantial and shock-enhanced magnetic fields in that tail also relaxed. This yields an offset between the spectral gradients of the two tails in the low frequency $\alpha$~map. Furthermore the few tail-aligned filaments that are still radio luminous from the shock activation of the $-z$ tail have largely dispersed and are by now weakened and more randomly oriented. High frequency emission in the $-z$~tail is largely absent aside from a few regions near the  AGN of relatively fresh and highly magnetized plasma. These outcomes are evident in figure~\ref{fig:M2obs2}, which illustrates the synchrotron properties at $t \approx 617$~Myr, just as the shock had completely passed through both tails.

\subsection{Synthetic Observations of the  Mach 3 Shock Encounter} \label{sec:M3obs}
Qualitatively, the synchrotron histories of the higher Mach number shock encounters, \textbf{M3} and \textbf{M4} are similar in many respects to the \textbf{M2} history just described. The principal distinctions come, as one might expect, from faster propagation speeds that shorten the duration of the encounters, while enhancing post shock turbulence, and stronger shock compressions in the higher Mach number shocks. We focus first on comparisons between the \textbf{M3} and \textbf{M2} encounters. 

\begin{figure}[ht]
\centering
\includegraphics[width=\textwidth]{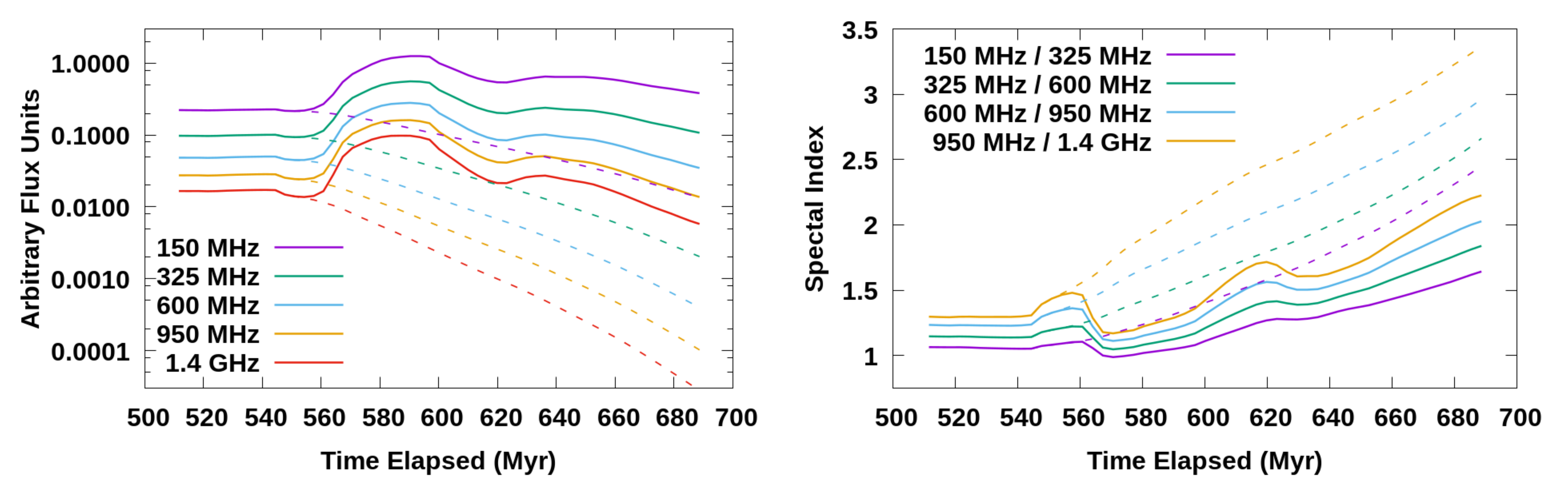}
\caption{Same as figure~\ref{fig:M2SpcEvo} for \textbf{M3} and \textbf{N3} starting at $t\approx 508$~Myr. The AGN jet source turns off at $t \approx 547$ Myr.} \label{fig:M3SpcEvo}
\end{figure}

The primary variances are obvious in a comparison between figures \ref{fig:M2SpcEvo} and \ref{fig:M3SpcEvo}, which show integrated fluxes and spectra over time. Note first that the full encounter lasted roughly $150$ Myr in the \textbf{M2} simulation, whereas the duration was only $\sim 90$ Myr for the \textbf{M3} encounter. Similarly, the peak flux enhancements during the \textbf{M3} encounter are roughly an order of magnitude relative to pre shock levels, compared to a factor $\sim 2$ in the \textbf{M2} case. Recall that, so long as the AGN jets remained active, the integrated fluxes and spectra of the unshocked NAT were almost stationary in time. During the \textbf{M2} event integrated spectral indices in the radio band generally increased over time once the AGN source switched off during the encounter. This was a manifestation of the predominance of radiative cooling of the CRe over reacceleration, plus a near absence of DSA, so that adiabatic compression and magnetic field amplification dominated synchrotron enhancements. By contrast, in the \textbf{M3} encounter there is an obvious spectral flattening over the entire band, revealing a significant role for DSA in substantial volumes.
\begin{figure}[ht]
\centering
\includegraphics[width=\textwidth]{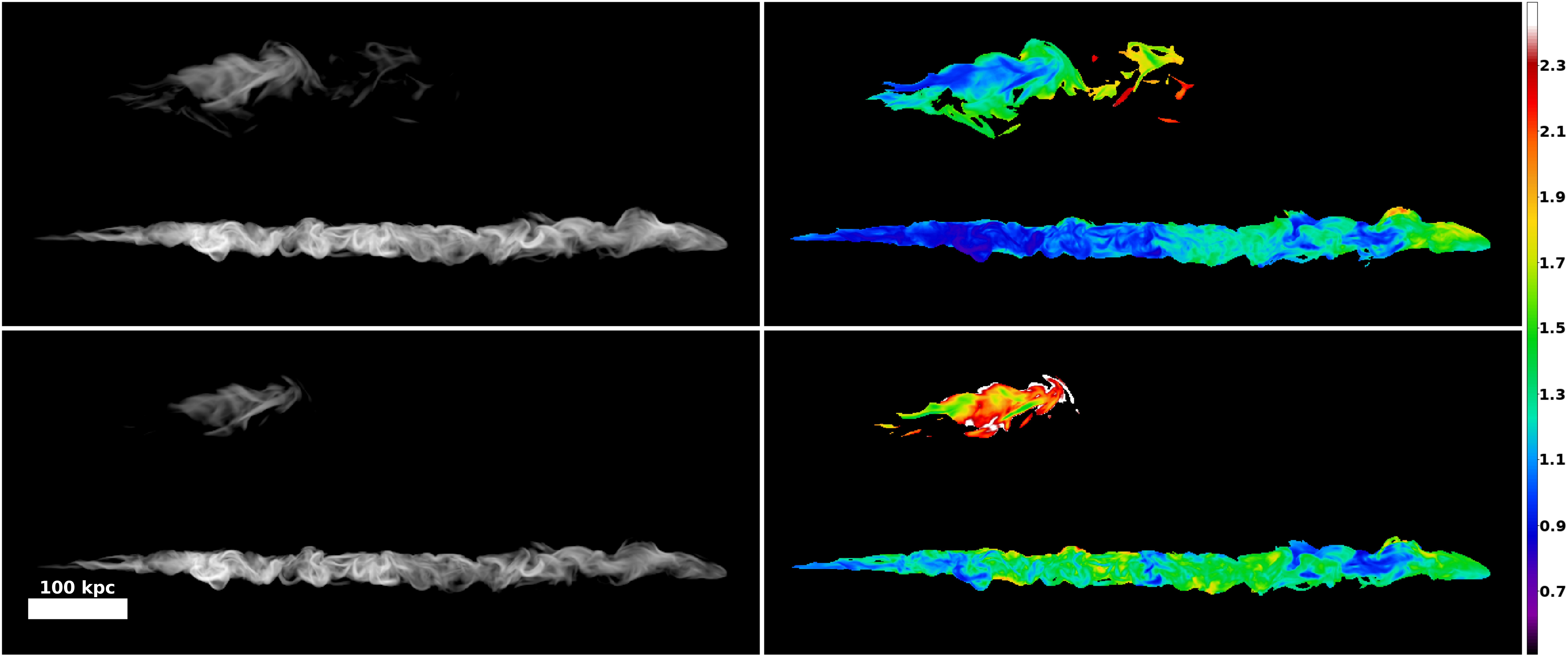}
\caption{Same as figure~\ref{fig:M2obs1} for \textbf{M3} at $t \approx 584$~Myr.} \label{fig:M3obs1}
\end{figure}

\begin{figure}[ht]
\centering
\includegraphics[width=\textwidth]{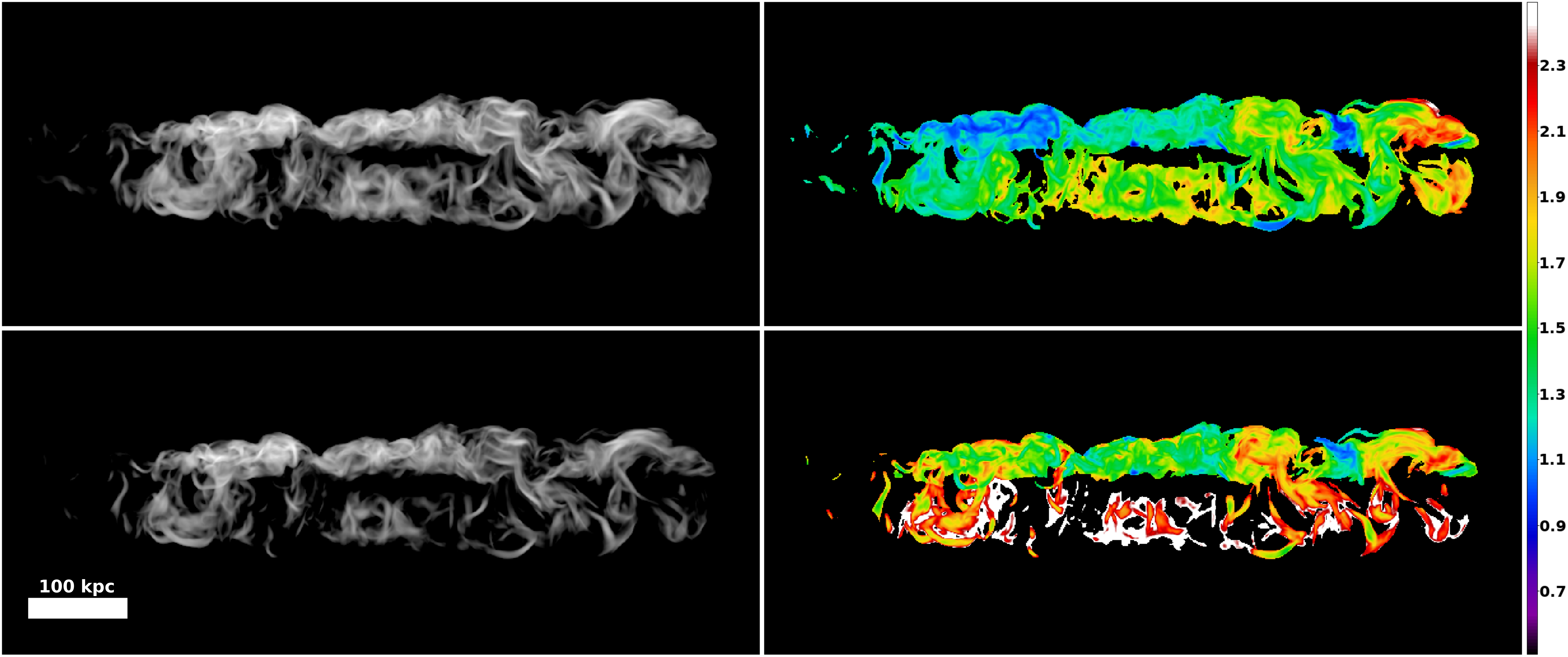}
\caption{Same as figure~\ref{fig:M2obs1} for \textbf{M3} at $t \approx 643$~Myr.} \label{fig:M3obs2}
\end{figure}

The dynamical differences and their consequences are also evident comparing figure  \ref{fig:M3obs1} to figure \ref{fig:M2obs1} and figure \ref{fig:M3obs2} to figure \ref{fig:M2obs2}. At the times shown in figures \ref{fig:M2obs1} and \ref{fig:M3obs1} the shocks have completed passage through the lower ($-z$) tail, but have not yet encountered the second tail. By contrast in figures \ref{fig:M2obs2} and \ref{fig:M3obs2} the shocks have completed their passage through both tails. It is obvious in figure \ref{fig:M3obs1} that the shocked tail is narrower than the analogous shocked tail in figure \ref{fig:M2obs1}. The lower tail in figure \ref{fig:M3obs1} can be divided left/right into two regions across a boundary roughly corresponding to the end of the  visible pre shock tail. As noted above, pre shock, significant aging led the older, downwind portions of the NAT to become essentially invisible. Now pretty much the full shocked tail is visible. Post shock, the lower energy CRe in the younger portions of the tail have primarily re-energized by adiabatic compression (Their spectra were not strongly aged before shock passage.). The post shock spectrum exhibits $\alpha_{950,1400} \la 1.5$. There is, on the other hand, a subtle spectral gradient extending roughly to the tail partition mentioned above. Downwind from there, where pre shock spectral aging was strong, post shock spectra, with some variations, commonly are characterized by $\alpha_{950,1400} \sim 1.3 - 1.5$, consistent with DSA from shocks with $\mathcal{M} \sim 1.7 - 1.9$. The principal exception to these spectral slopes comes in a region near the right end of the tail, where $\alpha \sim 0.9$ across the radio band, which would suggest DSA at a shock with $\mathcal{M} \sim 2.4$. This emission comes, in fact, from the same dense clump mentioned in \S \ref{sec:ShkVort}. While it is not exceptionally radio bright, its high mass density, dominated by ICM episodically entrained in the NAT tail, strengthened the shock in comparison to the neighboring tail. Such occurrences, point again to the post-shock consequences of the heterogeneous character of the NAT before shock impact.

Figure \ref{fig:M3obs2}, which presents images of the \textbf{M3} encounter after the shock has passed through both NAT tails, illustrates some additional influences of shock Mach number. In particular, the distortions imprinted on the shock during its passage through the lower tail have substantially greater influence on its modification of the second tail than was the case for the \textbf{M2} encounter. In particular, the reduced characteristic strength of the shock has significantly reduced the emission enhancements in the upper tail. For instance, DSA in the upper tail is mostly relatively unimportant outside of regions with densities well above the tail average. Consequently, by the time shown, the integrated flux at 150 MHz is only about half that when the shock finished passage through the first tail. Spectral steepening following termination of AGN inputs are well underway.

\subsection{Synthetic Observations of the  Mach 4 Shock Encounter} \label{sec:M4obs}

\begin{figure}[ht]
\centering
\includegraphics[width=\textwidth]{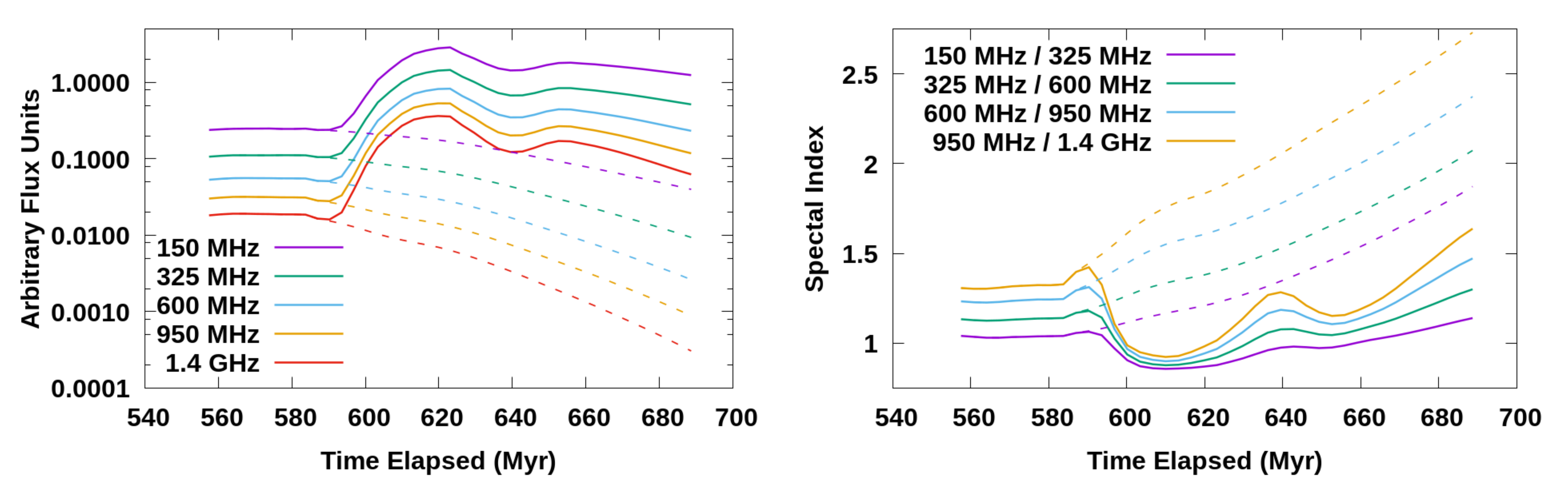}
\caption{Same as figure~\ref{fig:M2SpcEvo} for \textbf{M4} and \textbf{N4} starting at $t\approx 558$~Myr. The AGN jet source turns off at $t \approx 586$ Myr.} \label{fig:M4SpcEvo}
\end{figure}
The \textbf{M4} simulation features our strongest, and therefore fastest moving shock. The total encounter time with the NAT spans only about $65$ Myr, less than half the time associated with the \textbf{M2} encounter. However, magnetic fields are enhanced enough in portions of the tails that cooling of newly accelerated material does still play an important role. DSA during the shock passage through the $-z$ tail is broadly important and, as shown in figure \ref{fig:M4SpcEvo}, quickly flattens the integrated spectrum to $\alpha \sim 0.9$, corresponding to DSA from a shock with $\mathcal{M} \sim 2.4$. At the same time integrated spectral curvature is almost eliminated (again visible in figure \ref{fig:M4SpcEvo}). Post shock turbulent amplification of the magnetic fields in the lower tail continues to enhance the synchrotron brightness past when the shock completes its passage through that tail. By the time the shock begins to encounter the upper, $+z$ tail, around $t \approx 623$ Myr, the fluxes have peaked and the integrated spectrum shows signs of steepening at the higher frequencies we model. The peak fluxes are enhanced a bit more than an order of magnitude above the pre-shock levels.

\begin{figure}[ht]
\centering
\includegraphics[width=\textwidth]{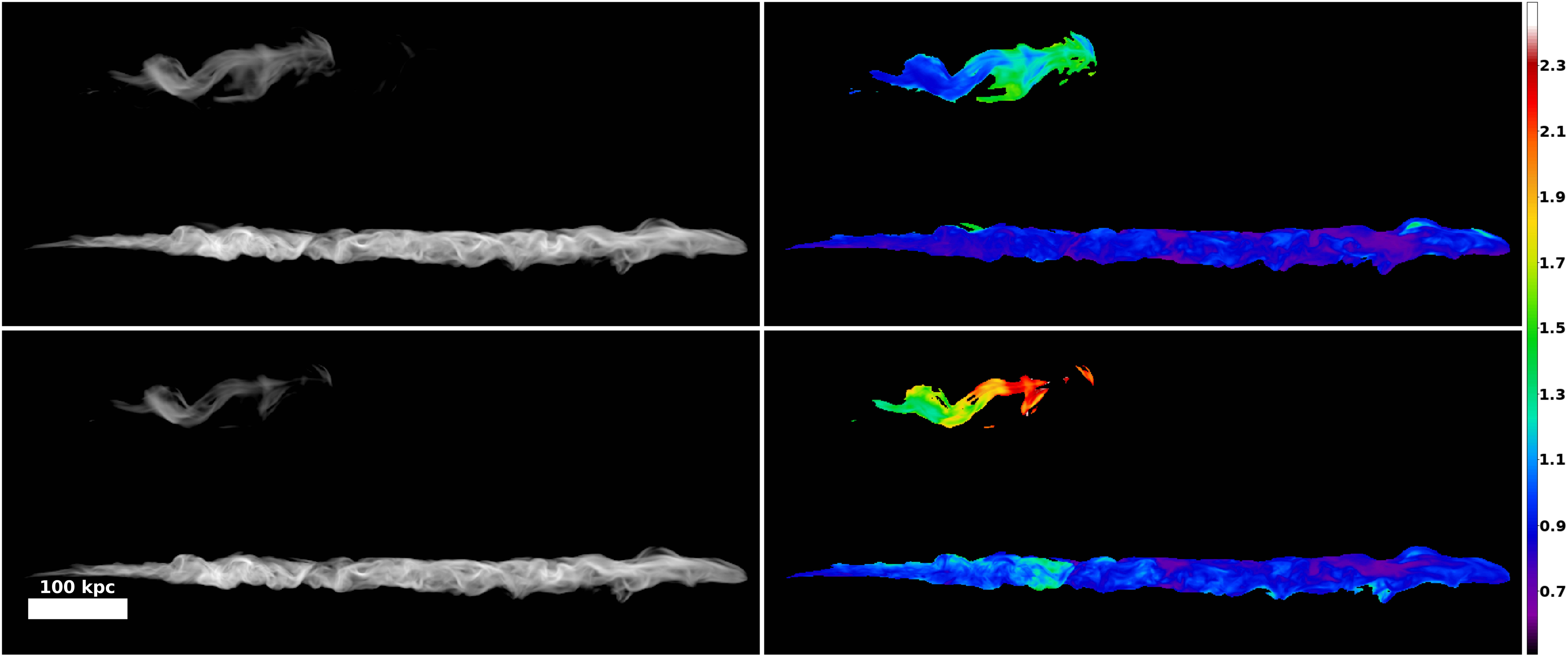}
\caption{Same as figure~\ref{fig:M2obs1} for \textbf{M4} at $t \approx 610$~Myr.} \label{fig:M4obs1}
\end{figure}
At the time of figure~\ref{fig:M4obs1}, the shock has just completed passage through the lower tail and compressed it to a maximum thickness of $\sim 25$~kpc, compared to a pre shock thickness roughly three times this. The brightest spots in the disruption zone dominate the $I_\nu$ images. Magnetic fields approach $10\muG$ in associated filaments, while intensities throughout the rest of the tail are fairly consistent averaging about half the brightness of the max intensity in both $I_{\nu}$ images. Again, low densities in the NAT near its source  lead to weak shocks in those regions. This ensures that the nominal DSA slope associated with the shock is steeper than the local CRe populations, so that adiabatic compression dominates DSA in those regions. Further downwind in the tail DSA does become the dominant particle acceleration process, leading to quasi power law emission spectra across the radio band. Spectral indices lie in the range $\alpha \sim 0.7$~-~0.9 ($\mathcal{M}\approx2.4$~-~3.3), with the variations mostly tracking density variations. 

\begin{figure}[ht]
\centering
\includegraphics[width=\textwidth]{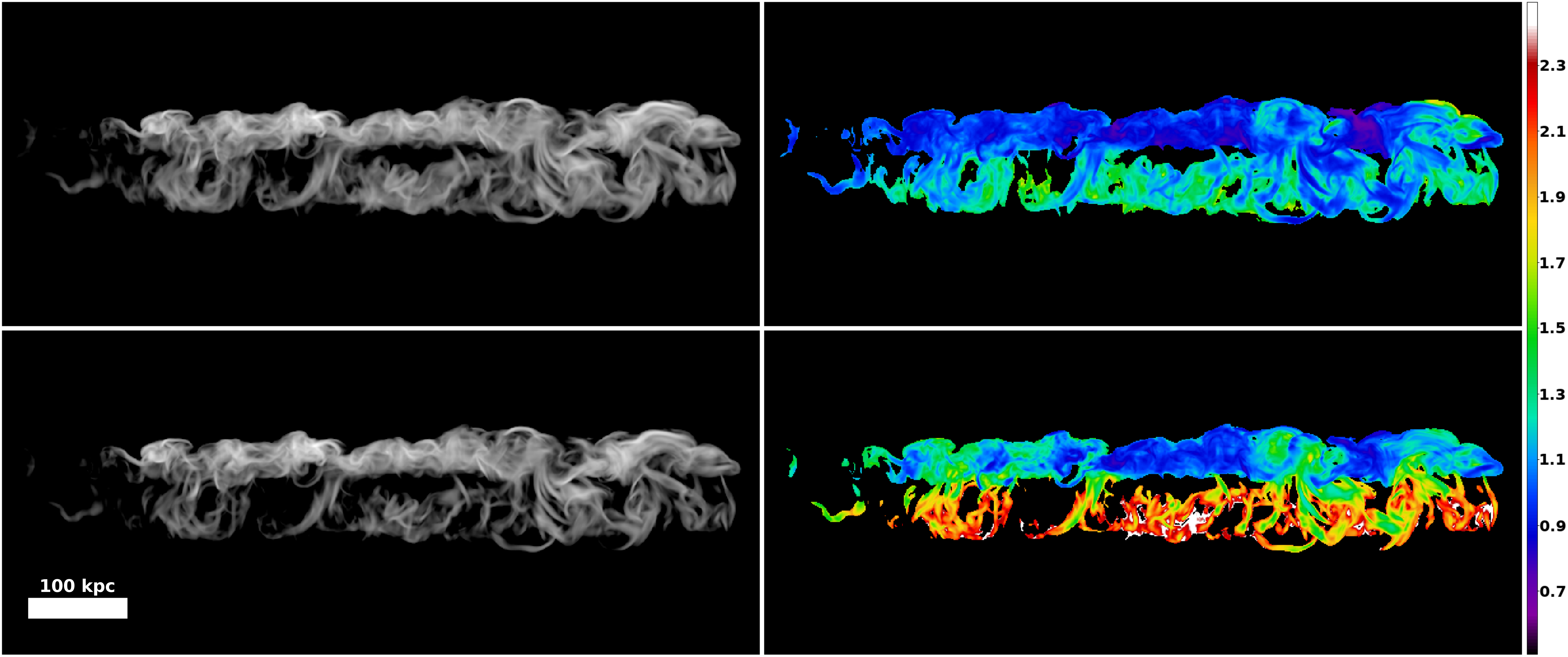}
\caption{Same as figure~\ref{fig:M2obs1} for \textbf{M4} at $t \approx 659$~Myr.} \label{fig:M4obs2}
\end{figure}
After shock passage, the integrated fluxes increase and then suddenly decrease, tracking the same magnetic amplification and diffusion seen in \textbf{M3} and \textbf{M2}, but on a faster timescale (compare figure~\ref{fig:M4SpcEvo} and figure~\ref{fig:M3SpcEvo}). The shock begins to run through the $+z$~tail around $t\sim640$~Myr, when radiative cooling and magnetic field decay have increased the spectral curvature such that $\Delta \alpha \approx 0.31$ over the span of our reference frequencies. By $t \approx 653$~Myr the shock has passed through the tail and this drops to $\Delta \alpha \approx 0.18$. In figure~\ref{fig:M4obs2} a few Myr after passing through the whole structure, the compressed $+z$~tail is nearly as narrow as the $-z$~tail is in figure~\ref{fig:M4obs1}, but is not as straight and ordered on account of the irregular distorted nature of the shock after passing through the first tail. The patches with highest intensity at 150~MHz are about 80\% as bright and at 950~MHz are about 70\% as bright as those in figure~\ref{fig:M4obs1}. The character of the $\alpha$~maps is largely very similar those in figure~\ref{fig:M3obs2} but with flatter spectra, as a result of the stronger shock and shorter cooling time for the CRe prior to reacceleration.

Figure \ref{fig:M4obs2} presents synchrotron images and spectra from the \textbf{M4} encounter just after the shock has finished crossing both NAT tails. The most obvious distinction here in comparison to the weaker shocks is that the dynamics has virtually merged the two tails into one, with the more recently shocked tail exhibiting a flatter spectrum. Superficially, the images resemble a scenario in which a plane shock upward from this view has reaccelerated a fossil CRe population embedded in the ICM.  On the other hand, a closer look reveals several features contrary to that scenario. For one, the consequences of the highly heterogeneous character of the pre shock NAT that has been so evident in our discussions made the brightness and spectral distributions of the synchrotron emissions highly nonuniform. We shall see in the following subsection, that these nonuniformities are at least as prominent in the radio polarization properties of the shocked NAT. In addition to these local inhomogeneities, there is a clear spectral discontinuity separating the two tails, at least from this viewing perspective. That results from the time offset between shock passage between the two tails. Nominally, a plane shock running over a simple, homogeneous CRe population would lead to a smoothly aged population. 

\subsection{Synthetic Radio Polarimetry of the Shocked NAT}\label{sec:polarimetry}

\begin{figure}[ht]
    \centering
    \includegraphics[width=\textwidth]{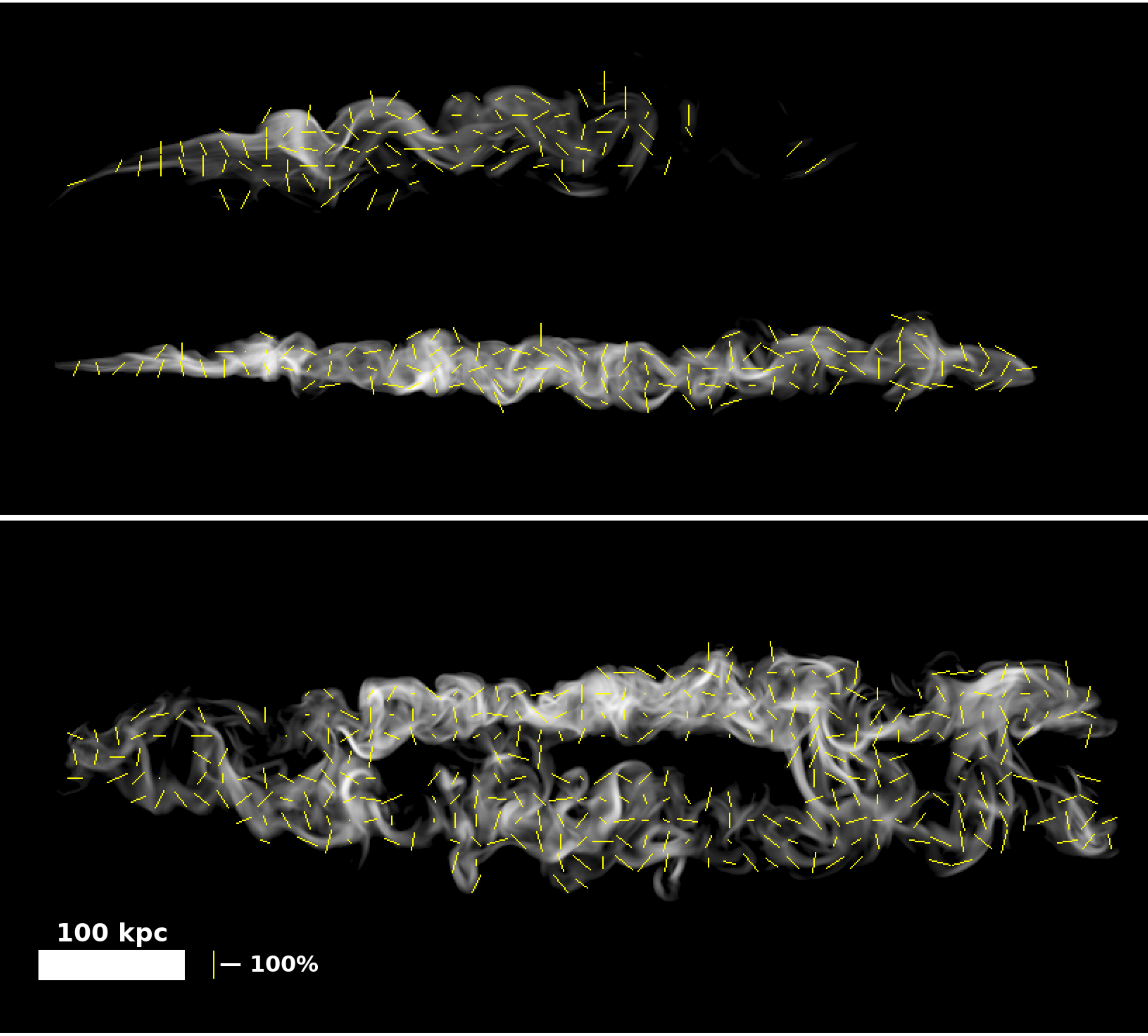}
\caption{Images from \textbf{M2} of $I_{\nu,150}$ with polarization vectors overlaid. Top: $t \approx 525$~Myr. Bottom: $t \approx 617$~Myr. Lines show orientation of E vectors at regularly spaced pixels. Their length is proportional to fractional polarization.}
    \label{fig:Pol2}
\end{figure}

We also used CRe distributions and magnetic field distributions to generate synthetic observations of the polarized radio emission from our shocked NAT, ignoring the effects of Faraday rotation\footnote{Rotation measures through the tails are of order unity or less for our parameters.}.  Figure~\ref{fig:Pol2} displays the polarization properties of 150~MHz emission from the \textbf{M2} run at the same times as figures~\ref{fig:M2obs1} and~\ref{fig:M2obs2}. The top image shows how the NAT would appear after the shock has completed its encounter with the $-z$ tail, and demonstrates the similar character of the polarized emission between the shocked and as yet unshocked tails. Large parts of both tails are locally highly polarized (up to $\sim 75\%$) and have integrated fractional polarization of $\sim 50\%$, but with a largely heterogeneous distribution of orientations that track the tangled topology of the magnetic fields. On the left-hand side of each tail, the remains of the coherent portion of each jet are still visible; most of the polarization vectors here are preferentially aligned vertically, so orthogonal to the tail axis and the shock plane as a result of the poloidal fields that dominate the jets along most of their lengths. In downstream portions of the tails, many of the bright, magnetized filaments have polarized emission vectors roughly perpendicular to the extension of the brighter filaments, whose projected orientations are quite variable. The shock does appear to have an effect on the integrated orientation; the emission-weighted orientation of the unshocked tail is $\sim 53^{\circ}$ from vertical while the shocked tail is more aligned, only $\sim 13^{\circ}$ from vertical even while locally the orientations are highly variable.

The bottom image displays the polarization in the emission after the shock has finished passing through the second, $+z$ tail and when the $-z$ tail has evolved for $\sim 100$~Myr in the turbulent post-shock flow. Aside from the lack of the coherent polarized tail sections mentioned at the earlier time, the character of the polarized emission is mostly the same. The upstream coherent jet remnants have continued their flow downstream and largely accumulated in the tails around the disruption disruption zone; portions of the jet plasma have been left behind in the jet channels which have been distorted by post shock turbulence. Locally, the vector orientations throughout the tails are largely uncorrelated with each other, as the preexisting topology of the magnetic fields dominates over any effects of the shock compression. In aggregate, the variations do mostly average out, with the net orientation only $\sim 10^{\circ}$ from vertical. Similar analysis of polarized emission of our \textbf{M3} and \textbf{M4} runs (not pictured) is consistent with these results. 

In summary, then, while the polarization can be very large for individual lines of sight in the shocked NAT, the highly heterogeneous magnetic field structure of the pre shock tails and the additionally heterogeneous evolution of the shocked magnetic field structures lead to complex polarization distributions and not polarization distributions expected from a plane shock propagating through a homogeneous medium.

\section{These simulations in the Context of Cluster Radio Relics}\label{sec:relics}

The elongated morphologies of NAT RGs could provide a convenient prospective source of fossil CRe to resolve theoretical issues that arise in attempting to explain observations (and nonobservations) of radio relics at cluster shocks. This possibility was among the motivations for this work. Additionally, several examples do exist in the literature where active radio galaxies with tails seem pretty clearly to associate dynamically with highly elongated cluster radio structures that, based on their general properties and locations within clusters, can reasonably be interpreted as cluster radio relics. In some cases shocks are also clearly present \citep[\eg][]{shim15}. However, our simulations suggest the need for caution, especially with regard to any universal interpretation of an association between NATs and cluster radio relics. In particular, some inherent features the synchrotron emissions of our shocked NAT described in \S \ref{sec:SynObs} are in significant conflict with some of the observed properties often associated with radio relics as a class. An outline of the common characteristics of radio relic observations can be found in a recent review by \citet{vanw19}, for example. In simple terms, the key radio relic characteristics at issue would be those that suggest clean plane shock interaction with a homogeneous medium leading to simple radio brightness and spectral distributions, along with high linear polarization with the polarization (E) vectors uniformly orthogonal to the major axis of the relic; that is uniformly aligned with the projected shock normal.

The key observational discrepancies we find are a consequence of the heterogeneous nature of the tail plasma that develops inherently from the dynamics behind the formation of our simulated NAT prior to the shock passage. For instance, the radio surface brightness in each of our simulated shocked NATs is highly uneven and dominated by a few regions where magnetic fields in the NAT plasma had recently been strongly enhanced by disruption processes prior to shock passage. While a few relics do have clearly irregular radio brightness distributions (e.g., \citealt{shim15,vanw16}), some of the more remarkable relics have almost constant intensities along their $\gtrsim 1$~Mpc lengths, with simple transverse gradients (most notably the so-called ``Sausage" relic in CIZA J2242.8+5301; \citealt{vanw10}). 

Among the more theoretically challenging properties of an object like the Sausage is the almost constant power law spectral index, $\alpha \approx 0.6$, along virtually the full length of the outer edge of the relic (generally interpreted as a simple shock with $\mathcal{M} \approx 4.6$) as well as a strong, consistent cooling gradient towards the inside (presumably the post shock direction). Contrary to this, each of our shocked tails display a range of spectral indices as a result of pre-shock variations in density, effective CRe age and the complexity of the shock propagation through these heterogeneous tails. That outcome may be better produced by a model such as one suggested by \citet{kang15} based on a large-scale ``cloud'' of fossil CRe well-mixed within a homogeneous ICM that is overrun by a cluster-scale, uniform strength shock. 

An additional conflict between shocked NAT properties and ``textbook'' radio relic properties involves synchrotron polarization behaviors. While the individual line of sight intensities from our simulated tails are highly linearly polarized, consistent with real-world radio relics, the orientations of the radio polarized vector orientations of our shocked NATs are highly variable, because the magnetic field topologies within are complex.  This variability contrasts with radio relic polarization vectors, which seem to be remarkably well organized and point to magnetic fields strongly aligned with their shock fronts. While we do see some increase in the alignment of the magnetic fields with the incident shock front after shock passage, the highly twisted, random topology of the pre shock magnetic fields in the simulated tails persists and post shock turbulence works over time to undo any preferential alignment by the shock. Again, this difference between the shocked NAT and the "textbook" radio relic may point to an origin involving relaxed, pre shock density and magnetic field distributions. 

Are there physically reasonable and likely modifications to NAT formation dynamics from those involved in our simulations that would eliminate these tensions?  They do not seem to us to be obvious. On the face of it, one would seem to need modifications that allowed the NAT tails to be much more ordered and homogeneous, possibly with magnetic fields regularly aligned along the tails. On the other hand, the very dynamics that allows light AGN jets to be deflected by a dense cross wind leads to inherently unstable flows that should mix AGN and ICM plasmas in complex ways. Those are the very properties that led to the conflicting behaviors. We conclude for now that while shocked NATs may be candidate precursors to radio sources in clusters with some of the properties generally associated with radio relics, the relics so-produced are likely to exhibit significantly inhomogeneous structures. Very likely some alternative, significantly ``cleaner'' origin is required to explain such iconic radio relics as the Sausage.

Still, as noted above, some cluster radio sources whose locations and general morphology suggest possible association with cluster shocks as well as likely radio galaxy origins and that have been labeled ``radio relics'' \citep[\eg][]{bona14,shim15} might still be candidates for dynamical generation within the scenario outlined here when they do not require highly regular internal structures. In addition, versions of the so-called ``radio phoenix'' scenario, in which a ``dead'', tailed radio galaxy becomes re-energized by a cluster shock \citep[\eg][]{enss02,mand19} could produce ``radio-relic-like'' sources, again, provided the homogeneity issues encountered here are not problematic. Alternative geometry radio-phoenix-generating shock RG encounters have also recently been simulated in \citep[\eg][]{nolt19b}.  These kinds of RG-shock encounters are, in fact, likely, and quite possibly can lead to observable cluster radio sources. The next generations of high sensitivity radio surveys, especially at low frequencies, are likely to turn up multiple examples, that could be understood within this class.

\section{Summary} \label{sec:sum}

In this paper, we have presented a simulations study of the interaction between large-scale shock fronts with characteristic strengths of galaxy cluster merger shocks (Mach numbers 2-4) and elongated tails composed of warm, low-density plasma suggested by models of narrow-angle tail (NAT) radio galaxy (RG) formation. The simulated NAT contained magnetic fields and cosmic ray electrons (CRe) carried in supersonic, light plasma jets modeled after those associated with cluster AGN. The simulation scenario was idealized in order to allow more direct focus on general behaviors rather than to attempt to model any specific object. In that spirit, although the AGN jets were magnetized and carried a passive CRe population whose energy evolution was followed, the ambient ICM was homogeneous (except for the incident shock), as well as unmagnetized and without embedded CRe. The CRe transport accounted for adiabatic and radiative (synchrotron and inverse Compton) energy changes, as well as diffusive shock acceleration (DSA). Injection of thermal electrons at shocks to the CRe population was not included.

The NAT used in these simulations was previously simulated and reported in \cite{on19a}. Also to reduce complications, the ICM shock impacted the NAT transverse to the NAT long axis, and with the shock normal in the plane containing both tails. Consequently, the shock interacted sequentially with the two tails. In addition to dynamical analyses, we reported synthetic radio observations in order to begin to assess the observable outcomes from such encounters. Since the basic scenario considered here has been proposed as one way to generate so-called peripheral radio relics in galaxy clusters, we also briefly compared the properties of our simulated objects to some of the properties commonly attributed to such radio relics. \\

The principle physical insights from this investigation are:

\begin{enumerate}
\item Shocked NATs can have physical and radio morphologies reminiscent of radio relic sources, with virtually the full physical length of the NAT structure illuminated by the shock passage. There are, however, significant tensions between the properties of the simulated objects and those seen in many radio relics. The most obvious of these follow.

\item The inherently heterogeneous character of the plasma that makes up the pre shock tails means the strength and speed of an initially uniform, planar shock will vary strongly within the tails. In denser plasma and plasma with older CRe populations (expelled earlier from the AGN source) where the shocked radio spectral index is a function of the Mach number, this can lead to intensity, spectral and polarization properties varying strongly and in complex ways throughout the shocked tail. This conflicts with some observed radio relics with smooth brightness and spectral profiles and that exhibit linear polarization vectors consistently orthogonal to the major axis of the relic.  Moreover, as our NAT tails are almost entirely at lower densities than the surrounding ICM, shock strengths in the tails are mostly weaker than in the ICM. On the other hand, the uneven density and velocity structures of the tail plasma and the faster propagation speed relative to the ICM means the shock front is highly distorted after passing through the first tail of our NAT, causing certain portions of the shock to be significantly weaker prior to encountering the second tail and making the interaction with the second tail significantly less ordered than the first. (Our simulation geometry led to sequential, transverse shock-tail encounters.)

\item The morphologies of the tails and their relatively low densities means that our ideally aligned merger shocks drive coherent vortical motions within the shocked tails. These vortical motions have the effect of clearing out much of the midplane of the tail on long enough timelines, and, because of the inhomogeneity of the tails, driving turbulent amplification of the magnetic fields. Field strengths continue to grow after shock compression for multiple Myr. This ``turbulent dynamo'' eventually gives way to a turbulence driven diffusion process, causing the magnetic fields strengths to weaken and randomizing their orientations.

\item The radio synchrotron intensities of our shocked tails tend to be dominated by one or more regions corresponding to clumps of strong magnetic filaments in the unshocked tails that continue to stand out in the distribution of magnetic fields after the tails are shocked. In our NAT, the magnetic fields reach a maximum in what we dub the disruption zone, where the nominally coherent jet is disrupted and transitions into the tail plasma consisting of a broadly varying mix of AGN and ICM plasmas. The quasi-regular, episodic nature of major jet disruption events in the evolution of the NAT means relatively strong filaments are spaced out along the length of the tails. After disruption, subsequent random, unsteady turbulent motions drive diffusion of magnetic energy from stronger filaments to the rest of the tail plasma. This diffusion is certainly partly numerical, but also related to the character of turbulence in our simulated tails. The strongly irregular quality of the radio brightness in our shocked tails is in contrast to observations of many radio relics. The intensities in our synthetic observations are remarkably filamentary (tracing the magnetic field topology), which is consistent with high resolution observations of multiple relics \citep{vanw19}. As the shocked tails radiatively cool in the simulations, a small number of stronger filaments tend to stand out and persist as the rest of the tail fades to obscurity. 

\item The spectral properties in our shocked tails naturally depend on the the strength of the shock involved, but also on multiple, observationally difficult-to-distinguish properties of the tail and the CRe populations. In our \textbf{M2} (Mach 2 shock),  run, the strength of the shock is weak enough within the tails that the simple DSA slope is steeper than the compressed slope of the pre shock particle population almost everywhere. Consequently, the initial spectral aging pattern of the NAT is preserved in the rejuvenated radio emission with adjustments for compression and magnetic field amplification. In our stronger shock simulations DSA does take place in some tail locations. The local density of the tail plasma, the strength of the magnetic fields and the age of the CRe can determine whether the emission we see at a given frequency has spectral properties derived from DSA or the pre-shock particle population adjusted for adiabatic compression and magnetic field amplification.
\end{enumerate}

This research was supported at the University of Minnesota by NSF grant AST1714205 and through resources provided by the Minnesota Supercomputing Institute. CN was supported by an NSF Graduate Fellowship under Grant 00039202. TWJ gratefully acknowledges the hospitality of KITP at UCSB where part of this work was completed under support of NSF grant PHY1748958. We acknowledge and appreciate the help and inspiration provided by Larry Rudnick and Avery F. Garon while this work was underway.  We also thank an anonymous referee for a thorough reading of the manuscript and thoughtful comments that helped us improve the presentation.

\end{document}